\documentclass[a4paper,11pt]{article}
\usepackage{rotating}
\usepackage{jheppub} 
\usepackage[utf8]{inputenc}
\usepackage[T1]{fontenc} 
\usepackage{placeins} 
\usepackage{chngpage}
\usepackage{amsmath,amsfonts}
\usepackage{graphicx}
\usepackage{color}
\usepackage{xcolor}
\usepackage{relsize}
\usepackage{epstopdf}
\usepackage{hyperref}
\usepackage{mathrsfs}
\usepackage{ragged2e}
\usepackage{amssymb}
\usepackage{placeins}
\usepackage[normalem]{ ulem }
\usepackage{amsthm}
\usepackage{comment}
\usepackage{graphics}	
\usepackage{caption}
\usepackage{subcaption}
\usepackage{verbatim}
\usepackage{arydshln}
\usepackage{slashed}

\usepackage{tabularx,booktabs}
\usepackage{multicol}
\usepackage{array,multirow}
\usepackage{booktabs}
\usepackage{colortbl}
\usepackage{float}

\renewcommand{\to}{\rightarrow}

\renewcommand{\max}{\mathrm{max}}

\newcommand{\be}{\begin{equation}}
\newcommand{\ee}{\end{equation}}
\makeatletter
\gdef\@fpheader{}
\vspace*{0cm}
\makeatother

\begin{document}

\title{On the coupling of axion-like particles to the top quark}

\author[a]{Fabian Esser,}
\author[b,c]{Maeve Madigan,}
\author[a,d]{Veronica Sanz}
\author[c]{ and Maria Ubiali}
\affiliation[a]{Instituto de F\'isica Corpuscular (IFIC), Universidad de Valencia-CSIC, E-46980 Valencia, Spain}
\affiliation[b]{Institut für Theoretische Physik, Universität Heidelberg, Germany}
\affiliation[c]{DAMTP, University of Cambridge, Wilberforce Road, Cambridge CB3 0WA, UK }
\affiliation[d]{Department of Physics and Astronomy, University of Sussex, Brighton BN1 9QH, UK}

\emailAdd{esser@ific.uv.es}
\emailAdd{madigan@thphys.uni-heidelberg.de}
\emailAdd{veronica.sanz@uv.es}
\emailAdd{mu227@cam.ac.uk}

\date{\today}

\abstract{In this paper we explore the coupling of a light  axion-like particle (ALP) to top quarks. We use high-energy LHC probes, and  examine both the direct probe to this coupling in associated production of a top-pair with an ALP, and the indirect probe through loop-induced gluon fusion to an ALP leading to top pairs. Using the latest LHC Run II data, we provide the best limit on this coupling. We also compare these limits with those obtained from loop-induced couplings in diboson final states, finding that the $t\bar t$+MET channel is the best current handle on this coupling. }


\maketitle

\section{Introduction}
 
Originally motivated by the QCD CP problem~\cite{Peccei:1977hh,Peccei:1977ur,Wilczek:1977pj,Weinberg:1977ma}, nowadays axion-like particles, or ALPs, are hypothesised in many extensions of the Standard Model (SM), see e.g. Ref.~\cite{Choi:2020rgn} for a recent review. In particular, they would appear in scenarios with a global symmetry, spontaneously broken down by a new confining sector. Such breaking would lead to new light particles, Goldstone bosons, which would act as ALPs.  

Traditionally, most studies had focused on the coupling of ALPs to photons and electron-positron pairs, and their cosmological, astrophysical and detector signatures. In these cases, the ALP mass range where these searches are sensitive is rather limited, typically below the KeV-MeV range.  On the other hand, ALPs could couple to any SM particle, and lead to novel signatures at colliders in a broader mass range. 
 
 Indeed, studies of the ALP coupling to gluons~\cite{Mimasu:2014nea} and diboson pairs~\cite{Jaeckel:2015jla,Brivio:2017ije, Bauer:2017ris,Craig:2018kne} using collider probes, have shown that the ALP can be searched for at colliders with a large mass reach, not only through resonant signatures but also via non-resonant production of a light ALP~\cite{Gavela:2019cmq,No:2015bsn,Carra:2021ycg}.  

On the other hand, there are important gaps in the collider studies of the ALP coupling to fermions.  In theories with no new flavour sources, the ALP-fermion coupling  is proportional to the fermion mass, which  provides  a strong motivation to  study  the ALP coupling to tops via LHC probes. 

The ALP-top coupling has not been studied in detail, and this paper aims to cover this gap, and motivate further collider studies. Previously, in  Ref.~\cite{Brivio:2017ije}, a search for a bosonic ALP was discussed in $t\bar t$+MET but no limit was obtained, whereas in a more recent paper~\cite{Bonilla:2021ufe} limits on the ALP-top coupling were obtained via their loop correction to the electron-positron coupling~\cite{AxionLimits}, but only for sub-MeV masses and in non-collider probes. Future prospects for the ALP-top coupling are discussed in Refs.~\cite{Hosseini:2022tac,Behr:2023nch,Rygaard:2023vyo}.  In this paper we will focus on probing LHC production of ALPs in a wide mass range through the ALP-top coupling. 

The paper is organised as follows. In Sec.~\ref{sec:theory} we will provide the theoretical framework to describe the ALP couplings to fermions and vector bosons. In Sec.~\ref{sec:model} we motivate the coupling of an ALP to  tops with a scenario of non-minimal Composite Higgs  with right-handed compositeness. In Sec.~\ref{sec:direct} we discuss the reinterpretation of a search for supersymmetry in a final state with fully leptonic top pairs and missing energy, which will provide the most stringent limit on the ALP-top coupling. In Sec.~\ref{sec:indirect} we use the loop-induced couplings to vector bosons via the top coupling, and obtain  a new indirect limit on this coupling. In the same section we also reinterpret the existing limits from ALP-gauge searches in terms of the ALP-top coupling, and discuss the constraints on this coupling from flavour changing neutral currents and electroweak precision tests. 
  In Sec.~\ref{sec:validity} we discuss the validity range of our current results. In Sec.~\ref{sec:others}, we motivate other signatures from an ALP-top coupling. Finally, in Sec.~\ref{sec:concls} we summarise the limits on 
the ALP-top coupling obtained in this analysis of current LHC data and conclude. 

\section{Theoretical framework}
\label{sec:theory}
Axion-like particles (ALPs) are pseudo-Goldstone Bosons with CP odd properties. As such, their couplings to other states are restricted by a shift symmetry,
$a \to a + C$, where $C$ is a constant. This symmetry, often associated with the spontaneous breaking of a global symmetry, constrains the possible couplings to other species. 

We will work under the assumption that the ALP is associated with a heavy new scale $f_a \gg v$, where $v$ denotes the electroweak scale.  In this regime, we may take an effective field theory (EFT) approach to describe the ALP interactions with the SM fields.  The effective Lagrangian describing all possible leading CP-even couplings of the ALP to the SM fields is given by Eq.~\ref{eq:app-lagrangian} in Appendix~\ref{app:ALPtheory}.  In particular, the ALP couplings to third-generation quarks are given by
\be
\label{eq:top_coupling_su2}
 \mathcal{L} =   \frac{\partial_\mu a}{f_a} \,  ( c_{Q_3}  \bar Q^3_L \gamma^\mu Q^3_L +  c_{t_R} \bar t_R \gamma^\mu t_R + c_{b_R} \bar b_R \gamma^\mu b_R) \ ,
\ee
 where $c_{Q_{3}}$, $c_{t_{R}}$ and $c_{b_{R}}$ denote the Wilson coefficients and $Q_{L}^3 = (c_{t_{L}}, c_{b_{L}})^T$.
Note that this generic coupling to fermions preserves the SM SU(2)$_L$ structure.  
Eq.~\eqref{eq:top_coupling_su2} includes only flavour-diagonal couplings, and we assume the ALP does not generate additional flavour structures\footnote{See Refs.~\cite{Bonilla:2022qgm,Carmona:2022jid, Carmona:2021seb,Chala:2020wvs,Bauer:2019gfk,Bauer:2021mvw} for works exploring the consequence of an ALP with flavour structure.}.

The purpose of this study is to investigate the ALP coupling to the top quark.  In particular, we will consider the following axial coupling of the ALP to the top quark:
\begin{equation}
\label{eq:top_coupling}
{\cal L} = 
  c_t \,  \frac{\partial_\mu a}{2 f_a} \,  (\bar t \gamma^\mu \gamma^5 t)   \ ,
\end{equation}
where $c_t = c_{t_{R}} - c_{Q_{3}}$ and $t = (t_L, t_R)^T$.
Throughout this paper we will investigate the impact of this ALP-top coupling on LHC data, working in the simplified scenario in which all other couplings of the ALP to the SM fields are set to zero at tree-level.  In Sec.~\ref{sec:model} we will discuss further motivation for this choice of coupling.  By making use of the equations of motion, this interaction can 
be rewritten as 
\be
{\cal L} = 
-i \, c_t \,  \frac{m_t a}{f_a} \,  (\bar t  \gamma^5 t) \ .
\ee
Therefore, the coupling of the ALP to fermions borrows the same hierarchy as the fermion masses. Naturally, the strongest fermion-ALP coupling would be to the third generation and, in particular, to the top quark, further motivating the study of this coupling.

The coupling of the ALP to fermions induces loop-corrections to all other SM particles, as explicitly computed in Ref.~\cite{Bonilla:2021ufe}. 
In particular, the ALP coupling to top quarks, $c_t$, would induce loop corrections to the ALP couplings to vector bosons, which can then be probed at the LHC, and take the generic form   
\begin{equation}
\label{Lgaugemass}
\mathcal{L} \supset - \frac{a}{f_{a}} ( c_{\tilde{G}}  G_{\mu \nu}^{a} \tilde{G}^{a \mu \nu} + c_{\tilde{W}} W_{\mu \nu}^{a} \tilde{W}^{a \mu \nu}+  c_{\tilde{B}} B_{\mu \nu} \tilde{B}^{\mu \nu} ) \, .
\end{equation}
After electroweak symmetry breaking (EWSB) we can write these new couplings as
\begin{eqnarray}
\label{Lgaugemass2}
    {\cal L} \supset  - &\frac{a}{f_a}& ( c_{a \gamma\gamma} F_{\mu\nu}\tilde F^{\mu\nu} +  c_{a \gamma Z}  F_{\mu\nu}\tilde Z^{\mu\nu} +  c_{a Z Z}  Z_{\mu\nu}\tilde Z^{\mu\nu} \\ \nonumber
    & +&   c_{a W W}  W_{\mu\nu}\tilde W^{\mu\nu} + c_{a g g} G_{\mu\nu}\tilde G^{\mu\nu}) \ ,
\end{eqnarray}
where
\begin{align}
    c_{a \gamma \gamma} &= s_w^2 c_{\tilde{W}} + c_w^2 c_{\tilde{B}}\\
    c_{a Z Z} &= c_w^2 c_{\tilde{W}} + s_w^2 c_{\tilde{B}}\\
    c_{a \gamma Z} &= 2 c_w s_w (c_{\tilde{W}} - c_{\tilde{B}})\\
    c_{aWW} &= c_{\tilde{W}}\\
    c_{agg} &= c_{\tilde{G}} \label{eq:Lgaugemass2} \, .
\end{align}
Note that in Ref.~\cite{Bonilla:2021ufe}, the couplings $g_{a XX}$ are related to the couplings in Eq.~\eqref{Lgaugemass2} by the relation 
\begin{equation}
\label{eq:ctog}
    c_{a XX}=\frac{f_a}{4} g_{a XX}. 
\end{equation}
Using the expressions given in Eqs.~\eqref{Lgaugemass2}-\eqref{eq:Lgaugemass2}, the coupling of the ALP to photons induced by a tree-level fermion coupling $c_f$ can be written
\begin{equation} 
\label{eq:photonloop}
   c_{a\gamma \gamma}^{\text{eff}} = \,    - \sum_{\text{f}}c_\text{f}  \,Q_\text{f}^2 N_C B_1 \left( \frac{4 m_\text{f}^2}{p^2} \right) \,,
\end{equation}
where $Q_f$ denotes the fermion charge, $N_C$ is the number of colours and $m_f$ denotes the fermion mass.  $B_1$ is a loop function that asymptotes to 1 at high $p^2$, but at low-energies vanishes, $B_1\to 0$.  At high energies, therefore, the $c_t$-induced ALP-photon coupling $c_{a \gamma \gamma}^{\text{eff}}$ is given by $c_{a \gamma \gamma}^{\text{eff}} = -(\alpha_{em}/3 \pi) c_t$.  Table~\ref{tab:recast} shows the explicit expressions of the high-energy limit of the loop-induced couplings to each of the SM vector bosons. Note that with this particular choice of ALP-top coupling, the $WW$ contribution vanishes.

\begin{table}[htb!]
\centering
\begin{tabular}{|c|c|}
\hline  Regime & Expression  \\ \hline\hline
    high-pT & $c^{\rm eff}_{a\gamma \gamma} = \,   - \frac{ \alpha_{\rm em}}{3 \pi}  c_\text{t} $  \\\hline
     high-pT & $c^{\rm eff}_{a\gamma Z} = \,     \frac{2 \alpha_{\rm em} s_w}{3 \pi c_w } c_\text{t}   $  \\\hline
      high-pT & $c^{\rm eff}_{a Z Z} = \,     -\frac{\alpha_{\rm em} s_w^2}{3 \pi c^2_w } c_\text{t}   $  \\\hline
      high-pT & $c^{\rm eff}_{a W^{+} W^{-}} = \,     0 $ \\ \hline 
       high-pT & $c^{\rm eff}_{a g g} = \,    - \frac{\alpha_s}{8 \pi} c_\text{t}   $  \\\hline
\end{tabular}
\caption{High-energy loop-induced SM gauge boson couplings to the ALP by the ALP-top coupling $c_t$.
\label{tab:recast} }
\end{table}

So far we have discussed the high-energy limit of the couplings loop-induced by the ALP-top coupling $c_t$. At  low energies, however, the loop-induced couplings can adopt different expressions. For example, the coupling of the ALP to photons induced by a tree-level fermion coupling, as given in Eq.~\eqref{eq:photonloop}, has the following low-$p^2$ limit:
\begin{equation} 
   c_{a\gamma \gamma}^{\rm eff} (p^2=0) \to 0 \, .
\end{equation}
Moreover, this dependence of the effective coupling on momentum suppresses the $c_t$-induced contribution to the decay of a light ALP to photons, in which case we need to evaluate the loop function at low $p^2=m_a^2$.  This is relevant for the decay channels of the ALP, and in particular for its stability at colliders, as discussed in Appendix~\ref{sec:stability}. 

On the other hand, the ALP couplings to other fermions induced by the ALP-top coupling $c_t$, adopt a different behaviour. For example, the effective ALP coupling to an electron-positron is as follows,
\begin{equation}
  c_{\rm eff}^e\simeq 2.48 \,c_t\, \alpha_{\rm em} \,\log{\Lambda^2/m_t^2} \, , 
  \label{ec:ceff}
\end{equation}
a contribution that does not vanish at any $p^2$ and that depends on the regularisation scale $\Lambda$, see Ref.~\cite{Bonilla:2021ufe} for more details.
Such a non-zero coupling to light fermions at low $p^2$ will play an important role when evaluating the collider stability of the ALP, see Appendix~\ref{sec:stability}.
 
\subsection{Model-building a coupling to the top}
\label{sec:model}

In this paper we  isolate the coupling of the ALP to top quarks, assuming this coupling is the leading interaction of the ALP with the SM, whereas interactions to other SM particles are derived from it by loop diagrams.  

Such an assumption is natural in many scenarios, for example in UV completions where the fermion mass hierarchy is addressed by preferentially coupling the top  to a mass-generation sector, where the ALP could reside.  

This scenario could be explicitly realised in models with extra-dimensions, particularly warped scenarios, or dual descriptions in Composite Higgs models with the so-called partial compositeness mechanism for mass generation. In this section we  provide an explicit example  of an  ALP which couples at leading order to top pairs based on partial compositeness. 

Let  us start with the composite Higgs set-up. The idea is that the Higgs doublet that we identify with the SM Higgs is actually a composite object, made of quarks and gluons from a new strong sector, beyond QCD, which confines and undergoes a phase transition near the electroweak scale. 

In general realisations of Composite Higgs models~\cite{Khosa:2021wsu}, the Higgs doublet is a subset of the pseudo-Goldstone bosons (pGBs) from the spontaneous breaking of a large global symmetry group, and one typically finds other pGBs coming alongside the Higgs doublet. Those additional pGBs can be singlets under the electroweak sector, and act as an axion-like particle. 

For example, in the {\it see-saw} Composite Higgs scenario~\cite{Sanz:2015sua, No:2015bsn}, one considers the sequential breaking of global symmetries
\begin{equation}
    SO(6) \to SO(5)\to SO(4) \simeq SU(2)\times SU(2) \ .
\end{equation}

At the end of the breaking chain, one of the $SU(2)$'s is weakly gauged to $SU(2)_L$ and the $T^3$ generator of the second is identified with $B-L$ and weakly gauged to represent hypercharge.
\begin{figure}[h!]
    \centering \includegraphics[scale=0.17]{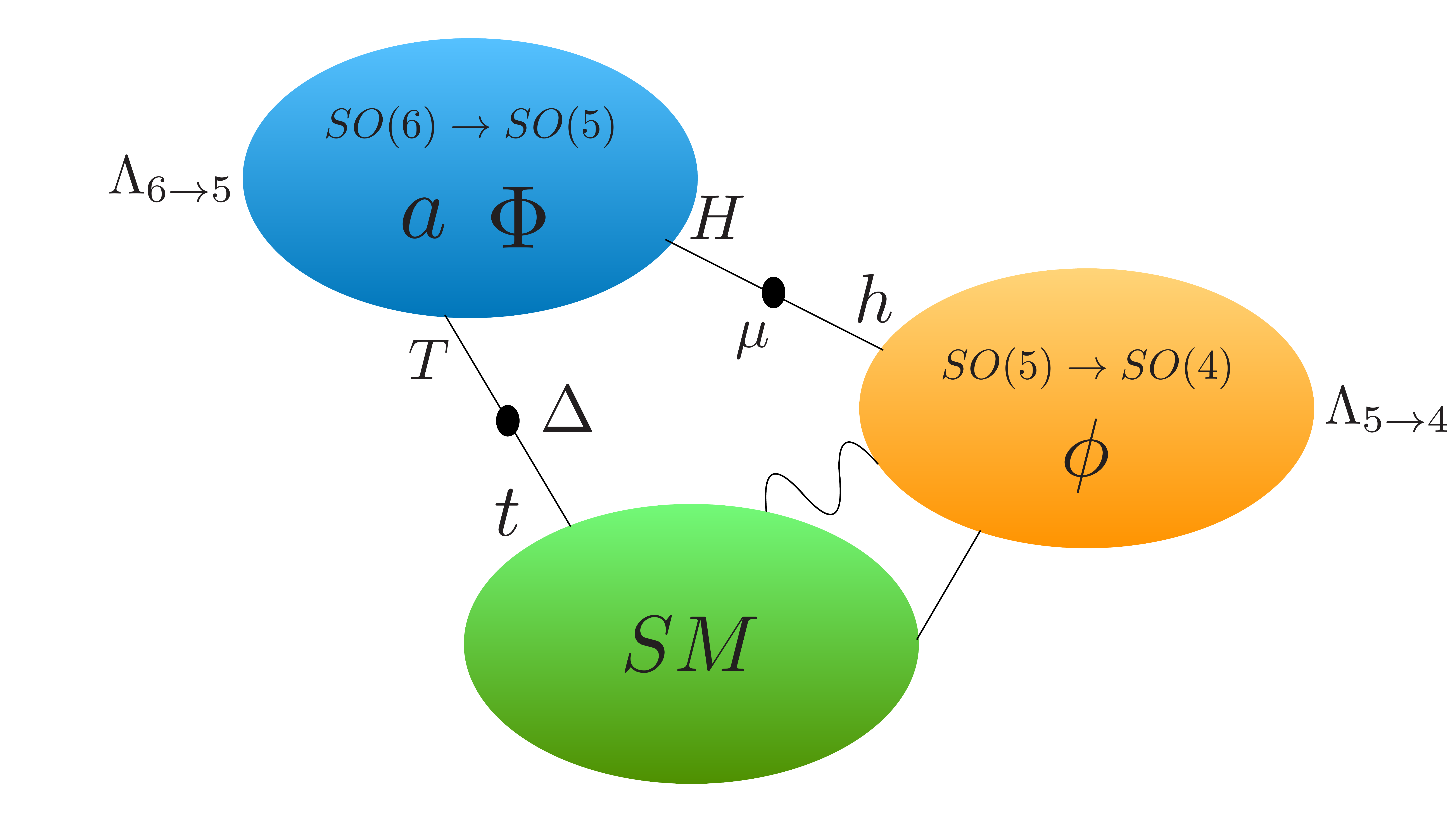}
    \caption{Schematic description of the see-saw Composite Higgs scenario and partial compositeness.}
    \label{fig:sketch}
\end{figure}
The first breaking  produces five pGB degrees of freedom, a doublet $\Phi$ and a singlet $a$. The second breaking leaves one doublet pGB  $\phi$. Strong interactions would produce a potential mixing of the two doublets, and the resulting mass eigenstates will end up being identified as the light Higgs $h$ and a new heavy Higgs $H$. In this scenario, the singlet degree of freedom $a$ would be naturally light and a candidate to represent the ALP we discuss in this paper. The scale $f_a$ would be linked to the first breaking $\Lambda_{6\to5}$ and the mass of the ALP would be zero, up to explicit breaking of the global symmetry.

In Fig.~\ref{fig:sketch} we show a sketch of the breaking and degrees of freedom.

Moreover, in these scenarios the mechanism responsible for triggering EWSB involves the presence of new fermionic composites, often called top-partners $T$, which in their simplest form are colored vector-like fermions with mass $m_T$ linked to the composite breaking scale $\Lambda_{6\to5}$. These states would couple to other composite states of the $SO(6)\to SO(5)$ breaking, including the light ALP $a$. By symmetry arguments, the coupling of the fermion $T$ with the ALP would have the form shown in Eq.~\eqref{eq:top_coupling},
\begin{equation}
    {\cal L} \supset -c_T \frac{\partial_\mu a}{\Lambda_{6\to5}} (\bar T \gamma^\mu T) \ .
\end{equation}
These baryonic states $T$ could mix with the top, by a mechanism usually called  {\it partial compositeness}. In particular, the new fermionic sector could couple to right-handed tops through a mass mixing,
\begin{equation}
    - \Delta \, \bar t_R \,  T + h.c.
\end{equation}
which would then induce a leading coupling of the ALP to top pairs, proportional to $c_t\propto c_T \frac{\Delta_R^2}{m_T^2}$. In this scenario, the ALP would  have no  couplings to other SM particles at the same order, and the leading contributions would be precisely those that we consider here, suppressed couplings via top loops. 

The explicit example we have discussed in this section is meant to illustrate how in UV completions an ALP coupled preferentially to tops could be a  consequence of a new sector linked to the mass generation mechanism. This could be realised in other forms, e.g. by coupling the third generation to new vector-like fermions $T$ and $\tilde T$ as
\begin{equation}
    -\Delta_L \, \bar t_L \, T_R - \Delta_R \, \bar t_R \, \tilde T_L \ .
\end{equation}
Alternatively, one could  consider  fermion representations which would allow couplings of new fermions to the SM fermion doublet $(t_L, b_L)$, denoted with $c_{Q_{3}}$ in Eq.~\eqref{eq:top_coupling_su2}, which would then lead to associated signatures involving the bottom quarks.

\section{Direct constraints on the ALP-top coupling}
\label{sec:direct}

Direct constraints on the ALP-top coupling $c_t$ can be obtained from processes in which an ALP is produced in association with a $t\bar{t}$ pair. We will focus on the process $p p \rightarrow t \bar{t} + a$ with a subsequent leptonic decay of the tops, cf.\ Fig.\ \ref{fig:ttbar_leptonic_diagrams}(a),  and will reinterpret a Run II ATLAS search for top squarks in events with two leptons, $b$-jets and missing transverse energy at $\sqrt{s}=13$ TeV and $\mathcal{L} = 139 $ fb$^{-1}$~\cite{ATLAS:2021hza}. 
For this analysis we assume the ALP to be collider stable, implying that it will escape the detector as missing transverse energy. For a dicussion of the ALP's collider stability see Appendix \ref{sec:stability}.

\subsection{On the kinematics of ALP production in association with tops}
\label{sec:ALP_kinematics}

\begin{figure}[htb!]
     \centering
     \begin{subfigure}[b]{0.4\textwidth}
         \centering
         \includegraphics[width=\textwidth]{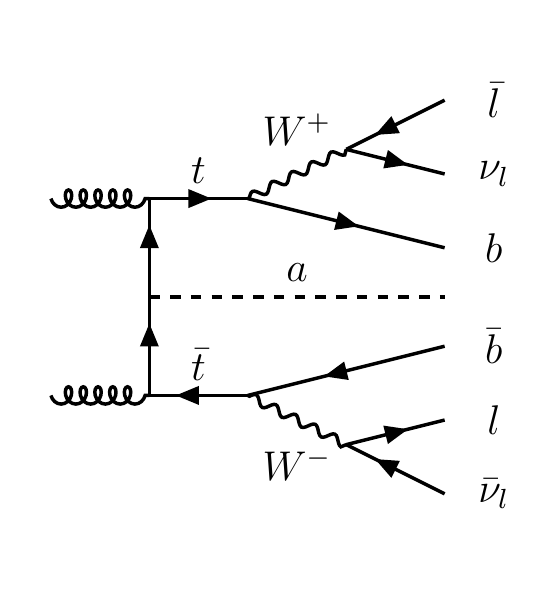}
         \caption{}
     \end{subfigure}
     \begin{subfigure}[b]{0.55\textwidth}
         \centering
         \includegraphics[width=\textwidth]{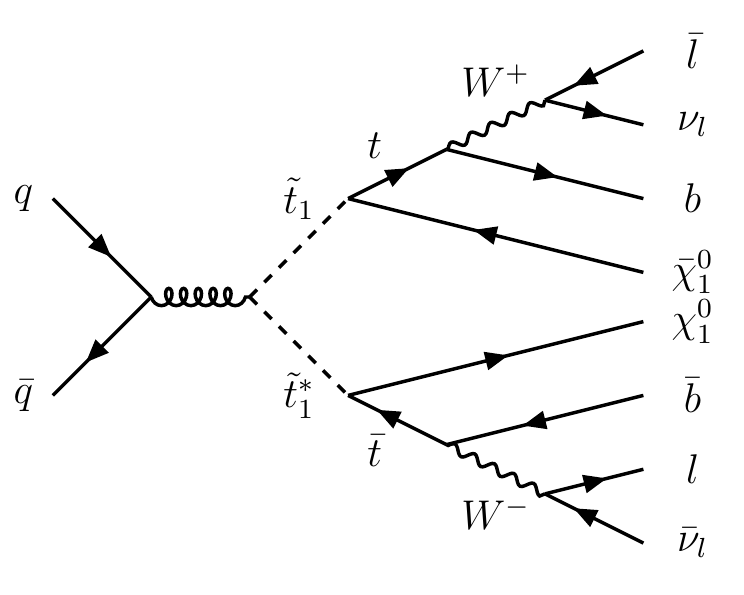}
         \caption{}
     \end{subfigure}
\caption{$t\bar{t}+a$ production from gluon fusion in the dileptonic decay channel (left), and an example for the SUSY benchmark $\tilde{t}{\tilde{t}}$ production from $q\bar{q}$ with leptonic top decay and neutralino decay (right). Note that $\tilde{t}{\tilde{t}}$ could also be produced via the s-channel process $gg \rightarrow g \rightarrow \tilde{t}{\tilde{t}}$ at the LHC.}
\label{fig:ttbar_leptonic_diagrams}
\end{figure}

Before presenting the details of the simulations and the results obtained from the direct search, we will investigate whether the search is sensitive enough to distinguish ALP signal events from the SM background. 
We will also compare the ALP kinematics with the standard supersymmetric (SUSY)  interpretations of this final state, where the top pairs and missing energy are produced via pair production of stop pairs,
 with a prompt decay into top quarks and neutralinos,
\begin{equation}
p p \rightarrow \tilde{t}_1 + \tilde{t}_1^{\dagger}, \quad \left(\tilde{t}_1 \rightarrow t + \chi_1^0 \right), \quad \left(\tilde{t}_1^{\dagger} \rightarrow \bar{t} + \bar{\chi}_1^0 \right),     
\end{equation}
see Fig.\ \ref{fig:ttbar_leptonic_diagrams} (b).
We will consider the top-quark pair to decay leptonically and the ALP to be collider stable. Therefore, the three scenarios (SM, ALP, SUSY) lead to the same final state topology of 2-leptons + 2 jets + missing transverse energy. 
The missing energy component is different in each scenario: neutrinos in the SM case, and  $a$ or $\tilde\chi^0$ particles  escaping detection for the ALP and SUSY signals.

To illustrate the differences in kinematics for these three cases, we will show two kinematic variables: first, the transverse momentum of the event $p_T^{\rm miss}$, and second,  an angular distribution, the distribution with the azimuthal angle between the sum of boosted momenta and the missing momentum,  $\Delta \Phi_{\rm boost}$,  which is defined as
\begin{equation}
    \Delta \Phi_{\rm boost} = \arccos \left(\frac{\vec{p}^{\rm miss}\cdot \vec{p}^{\rm boost}}{|\vec{p}^{\rm miss}| |\vec{p}^{\rm boost}|} \right)
\end{equation}
with $\vec{p}^{\rm boost} = \vec{p}^{\rm miss} + \vec{p}_{l_1} + \vec{p}_{l_2}$. 

\begin{figure}[h!]
    \centering
    \includegraphics[scale=0.8]{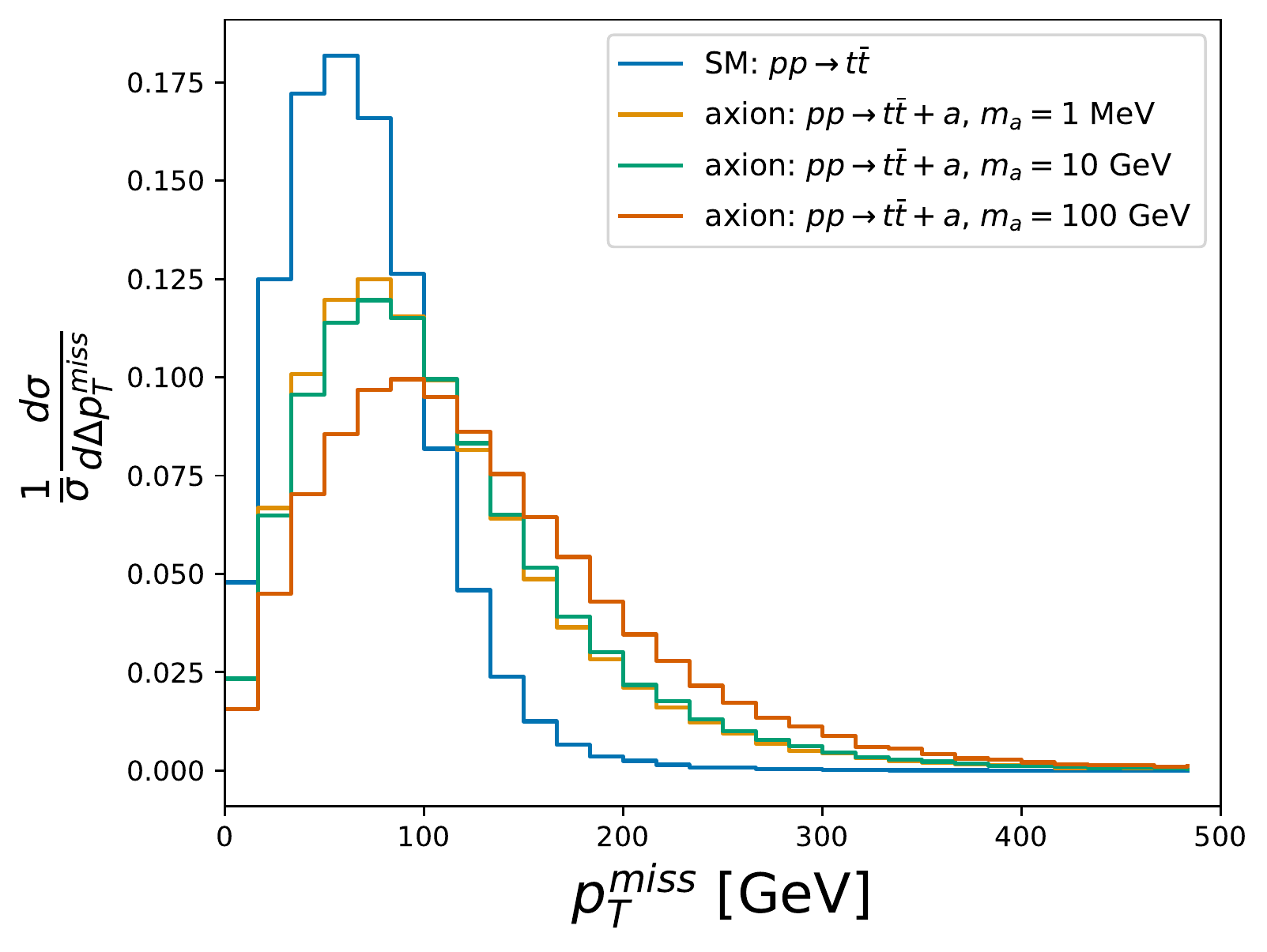}
    \caption{Normalised differential cross section (GeV$^{-1}$) in $p_T^{\rm miss}$ for the SM background and ALP signal in a fully leptonic decay. The signal events assume $c_t =1$ and $f_a = 1$ TeV. Curves are shown for $m_a = 1$ MeV, $m_a = 10$ GeV and $m_a = 100$ GeV.}
    \label{fig:ptmiss_distribution}
\end{figure}

\begin{figure}[htb!]
    \centering
    \includegraphics[scale=0.8]{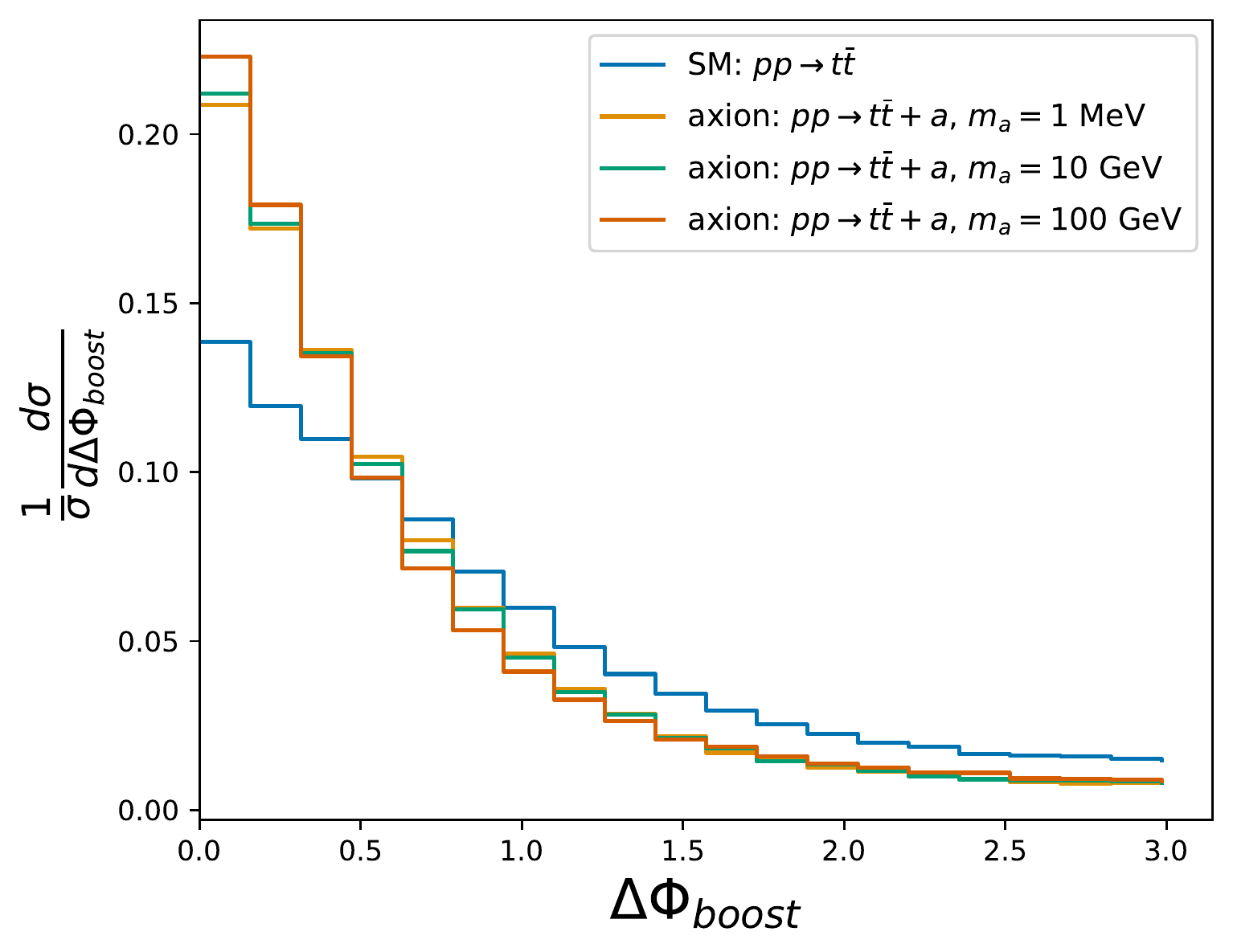}
    \caption{Normalised differential cross section in $\Delta \Phi_{\rm boost}$ for the SM background and ALP signal in a fully leptonic decay. The signal events assume $c_t =1$ and $f_a = 1$ TeV. Curves are shown for $m_a = 1$ MeV, $m_a = 10$ GeV and $m_a = 100$ GeV again.}
    \label{fig:DeltaPhiBoost_distribution}
\end{figure}

\begin{figure}[htb!]
    \centering
     \centering
     \begin{subfigure}[b]{0.49\textwidth}
         \centering
         \includegraphics[width=\textwidth]{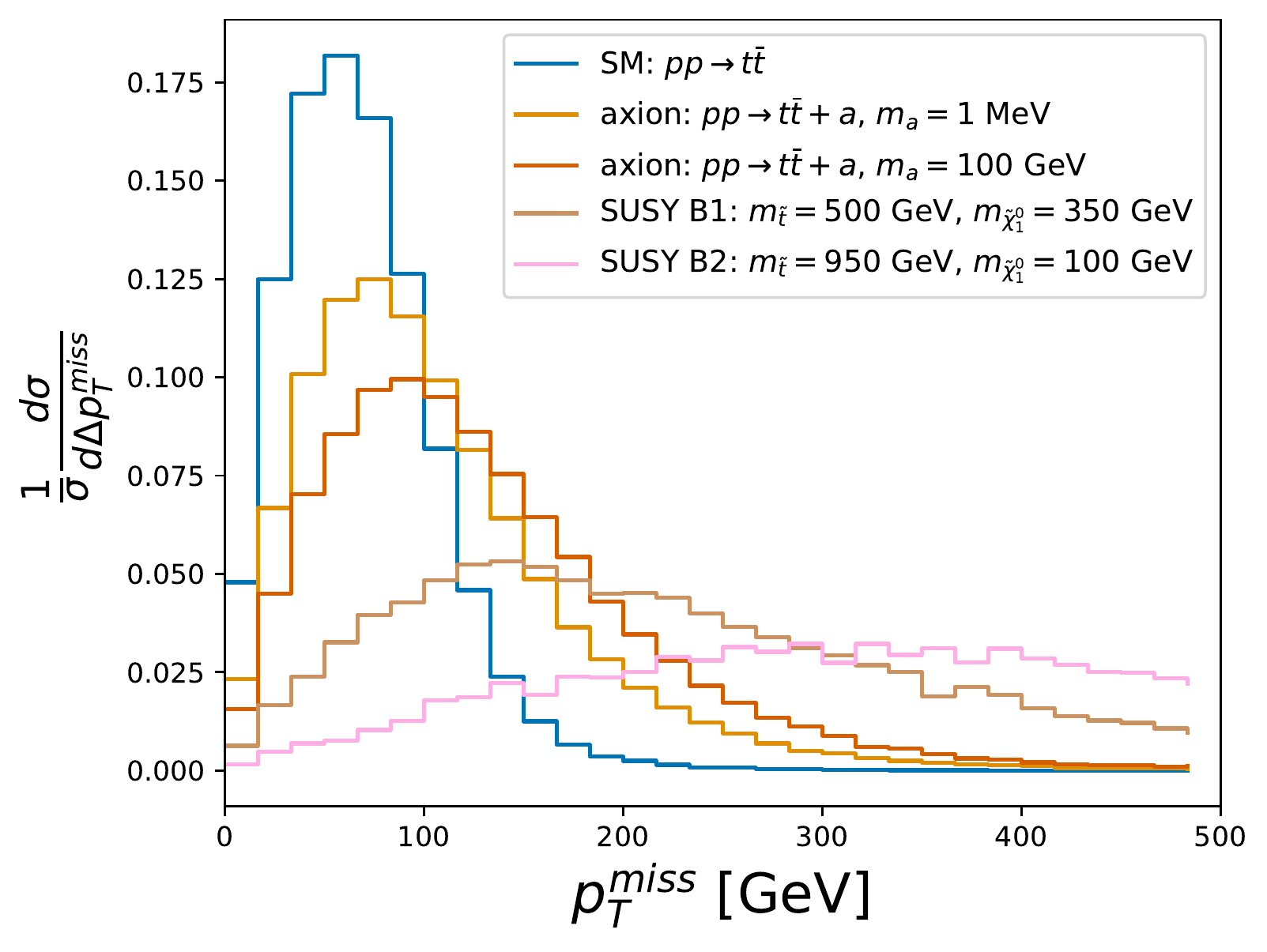}
         \caption{}
     \end{subfigure}
     \begin{subfigure}[b]{0.49\textwidth}
         \centering
         \includegraphics[width=\textwidth]{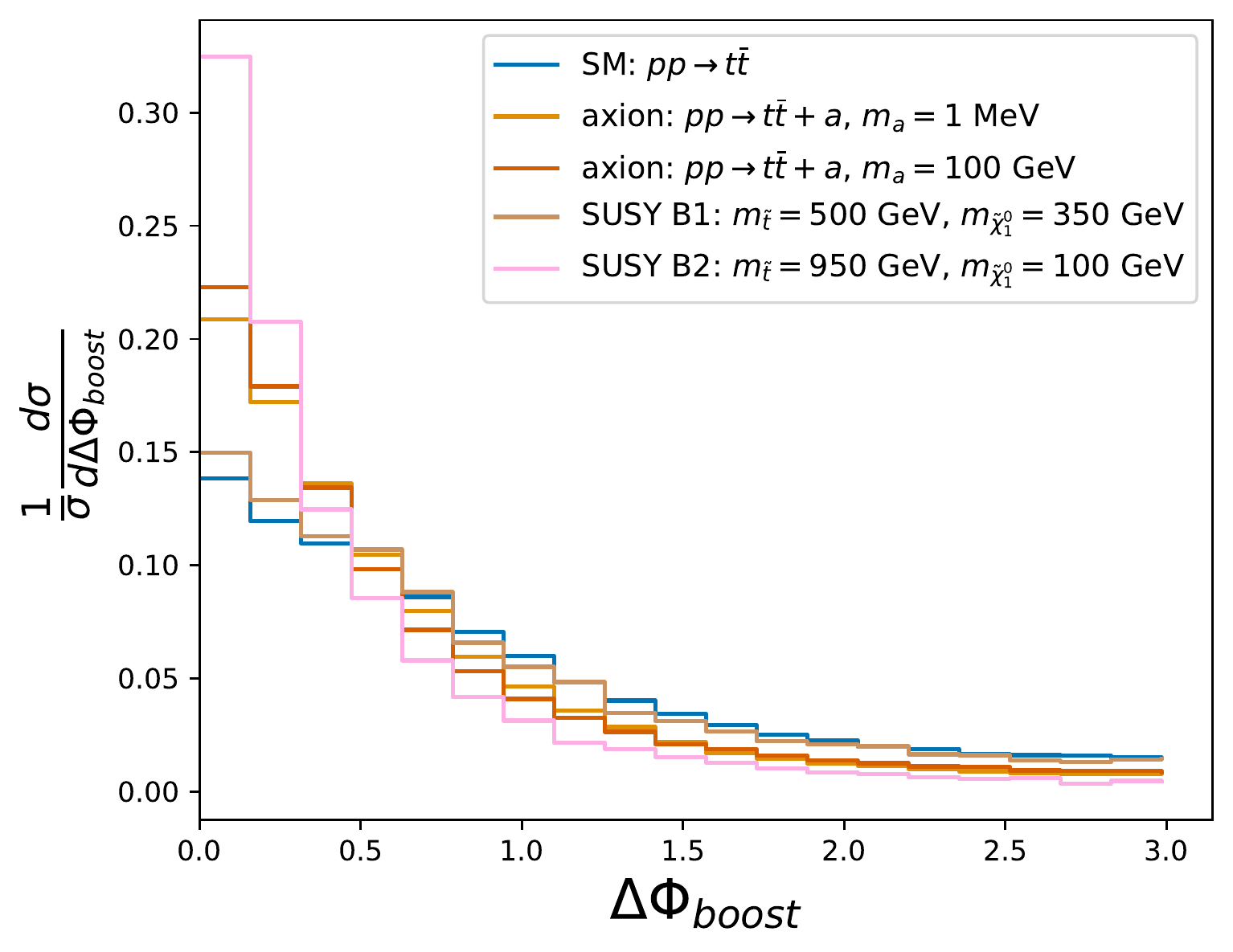}
    \caption{}
    \end{subfigure}
    \caption{Normalised differential cross sections in $p_T^{\rm miss}$ and $\Delta \Phi_{\rm boost}$ for the SM background, the ALP signal and two SUSY benchmark models in a fully leptonic decay. The ALP signal events assumes $c_t =1$ and $f_a = 1$ TeV.}  \label{fig:susy_distributions}
\end{figure}

Fig.~\ref{fig:ptmiss_distribution} shows the distribution with $p_T^{\rm miss}$ for the pure SM background process $pp \rightarrow t\bar{t}$ and ALP signal events with $m_a = 1$ MeV, $10$ GeV and $100$ GeV. 
It is clearly visible that for all considered ALP masses the event distribution for processes with a final-state ALP are boosted towards higher missing transverse momenta. While the ALP curves for $m_a = 1$ MeV and $10$ GeV are not distinguishable from each other, the distribution for $m_a = 100$ GeV is slightly shifted towards larger missing transverse momenta. We conclude that a reinterpretation of the SUSY search might display a good sensitivity to the ALP scenario due to being boosted with respect to the SM background. 

Fig.~\ref{fig:DeltaPhiBoost_distribution} depicts the distribution with $\Delta \Phi_{\rm boost}$, again for the pure SM background process and the same values of $m_a$ as above. In this case the ALP events feature a higher peak for small boost angles and a smaller tail towards higher values compared to the SM background process. The different ALP masses on the other hand do not alter much the shape of the distribution.

After discussing the kinematic differences between the ALP signal and the SM background distribution, we explore how the ALP signal compares to the SUSY signal, for which the ATLAS search was originally designed.
In Fig.~\ref{fig:susy_distributions} we show the distributions of $p_T^{\rm miss}$ (left panel) and $\Delta \Phi_{\rm boost}$ (right panel), respectively, for two SUSY benchmark scenarios.
The first SUSY benchmark scenario B1 assumes a light stop $m_{\tilde{t}_1} = 500$ GeV and an intermediate neutralino $m_{\chi_1^0} = 350$ GeV, while the second SUSY benchmark scenario features a heavy stop $m_{\tilde{t}_1} = 950$ GeV and a light neutralino $m_{\chi_1^0} = 100$ GeV. These two benchmarks are motivated by the current exclusion limits on stops from the search for fully-leptonic $t \bar t$+MET signatures~\cite{CMS:2020pyk}. 
For better visibility we do not show the curve for $m_a = 10$ GeV in Fig.~\ref{fig:susy_distributions} since, as observed in Figs.~\ref{fig:ptmiss_distribution} and ~\ref{fig:DeltaPhiBoost_distribution}, the distribution does not depend on the ALP mass, as long as the mass is smaller than around 100 GeV.  

The $p_T^{\rm miss}$ distribution shows that the SUSY spectra are shifted towards even larger missing transverse momenta than the ALP signals. This could be expected as the two heavy neutralinos both contribute to $p_T^{\rm miss}$. The shift is even larger for $B2$ although the neutralino is lighter, but in this case the neutralinos experience a stronger recoil effect due to the larger mass gap between the initial stop and the neutralino.
Looking at the difference between the distributions with $\Delta \Phi_{\rm boost}$, we note that SUSY models prefer smaller angles than those in the ALP models under scrutiny and an even more suppressed tail towards larger momenta. 
In summary, these distributions show that 
the ALP signal is boosted with respect to the SM background, and favours smaller values of $\Delta \Phi_{\rm boost}$ than the SM background. Although these differences are not as large as the difference between the SUSY benchmark models discussed and the SM background, we conclude that the phase space cuts defining the signal region of these SUSY benchmark models, listed in Table~\ref{tab:atlas_cuts}, will be well-suited to the search for a $p p \rightarrow t \bar{t} + a$ signal.

\subsection{ALP signal simulation and constraints on $c_t$ from the ATLAS search}
\label{sec:direct_constraints_ATLAS}
Next we will study the ALP signal $pp  \rightarrow t \bar{t} + a$ and reinterpret the ATLAS SUSY search in \cite{ATLAS:2021hza} in terms of this ALP signal.  
ATLAS provides a measurement of the $m_{T2}$ distribution in the 2-leptons + 2 jets + MET final state with different lepton flavours, using the LHC Run-II data with $\mathcal{L} = 139$
fb$^{-1}$ and $\sqrt{s}=13$ TeV, and a SM background estimate.
The stransverse mass $m_{T2}$ is defined by
\begin{equation}
m_{T2}(\vec{p}_{T,1}, \vec{p}_{T,2}, \vec{p}_T^{\rm miss}) = \min_{\vec{q}_{T,1} + \vec{q}_{T,2} = \vec{p}_T^{\rm miss}} \{ \max \left[ m_T(\vec{p}_{T,1}, \vec{q}_{T,1}), m_T(\vec{p}_{T,2}, \vec{q}_{T,2}) \right] \},
\end{equation}
i.e.\ we minimise over all distributions of the momenta of the neutrinos for fixed $p_T^{\rm miss}$ and take the maximum of the transverse masses for the two lepton-neutrino pairs, which are defined as
\begin{equation}
    m_T(\vec{p}_T, \vec{q}_T) = \sqrt{2 |\vec{p}_T||\vec{q}_T|(1- \cos(\Delta \Phi))} \, .
\end{equation}
$\Delta \Phi$ denotes the azimuthal angle between the two transverse momenta, and the particle masses are neglected. 

Instead of the full phase space, the analysis is performed in the fiducial region which is approximated by the phase space cuts listed in Table \ref{tab:atlas_cuts}.  The data are presented in 7 bins with increasing widths towards higher $m_{T2}$.  The bin edges are given by (100, 110, 120, 140, 160, 180, 220, 280) GeV.

\begin{table}[]
    \centering
    \begin{tabular}{c|c}
    \hline
    parameter & value \\
    \hline \hline
        $p_T$ leading lepton & $> 25$ GeV  \\
        $p_T$ subleading lepton & $> 20$ GeV \\ 
        \hline
        $m_{ll}$ & $ > 20$ GeV \\
        $m_{T2}(ll)$ & $> 110$ GeV \\
        $|m_Z - m_{ll}|$ & $ > 20$ GeV \\
        \hline
        $n_{\text{b-jets}}$ & $\ge 1$ \\
        $\Delta \Phi_{\rm boost}$ & $ < 1.5$ rad \\
        \hline
    \end{tabular}
    \caption{Phase space cuts defining the signal region in the ATLAS search for $t \bar{t}$ + MET~\cite{ATLAS:2021hza}.}
    \label{tab:atlas_cuts}
\end{table}
To simulate the ALP signal, we generate events for the process
\begin{equation}
    p p \rightarrow t \bar{t} + a,\, \left(t \rightarrow W^+ + b, W^+ \rightarrow l_1^+ \nu_{l_1} \right), \left(\bar{t} \rightarrow W^- + \bar{b}, W^- \rightarrow l_2^- \bar{\nu}_{l_2} \right) \, ,
\end{equation}
imposing that $l_1, l_2 \in \{e, \mu\}$ have different flavour, at LO
with {\tt MadGraph5\_aMC@NLO}~\cite{Alwall:2014hca}, utilizing {\tt NNPDF4.0}~\cite{NNPDF:2021njg} in the 4-flavour scheme. We employ the UFO model {\tt ALP\_linear\_UFO}~\cite{ALPUFO}, in which we set the ALP scale $f_a = 1$ TeV, fix the ALP mass to $m_a = 1$ MeV and -- since we are only interested in ALP-top couplings at tree-level -- set $c_{a\Phi} = 1$ and all other ALP couplings to zero.  See Appendix~\ref{app:ALPtheory} for more details on the relationship between $c_{a \Phi}$ and $c_t$, the use of the UFO model {\tt ALP\_linear\_UFO}.

The cut on $\Delta \Phi_{\rm boost}$ has been implemented in the {\tt MadGraph5} Fortran routines at the stage of event generation. The $m_{T2}$ cut is only applied at the analysis level as this is the variable in which we present the distribution.
We make use of the python package \texttt{mt2} to implement this cut and define the $m_{T2}$ bins~\cite{Lester:1999tx,Lester:2014yga}.
All the other cuts in Table \ref{tab:atlas_cuts} are implemented directly in the {\tt MadGraph} runcard.

\begin{figure}[htb!]
    \centering
    \includegraphics[scale=0.75]{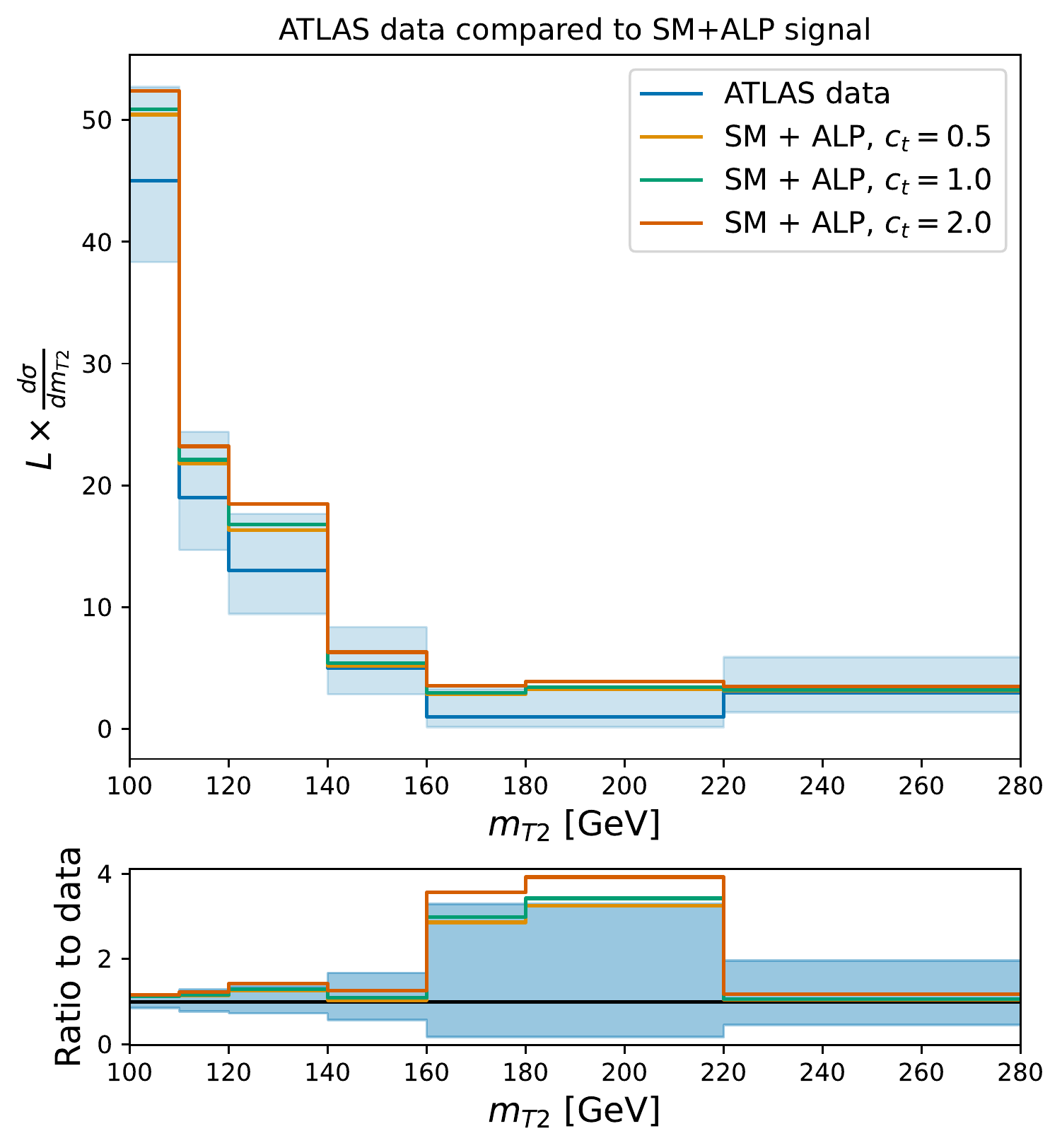}
    \caption{Comparison of the ALP signal events with SM predictions and ATLAS data in the $t\bar{t}+$MET final state as a function of $m_{T2}$, for different values of the coupling $c_{t}$.  The values $f_a=1$ TeV and $m_a = 1$ MeV are used to generate the ALP signals. The y-axis shows the number of events per binwidth (in GeV).}
    \label{fig:direct_ATLAS}
\end{figure}

There is one caveat to have in mind before we can compare the ALP signal to the SM background and the data:
since we generate the ALP signal in {\tt MadGraph} at leading order and we do not take into account higher order corrections, hadronisation and detector effects, we must calculate a normalisation factor between the full ATLAS simulation and our simulation, which we can then use to rescale the ALP signal.
We chose the process 
\begin{equation}
    pp \rightarrow t \bar{t}, \left(t \rightarrow W^+ + b, W^+ \rightarrow l_1^+ \nu_{l_1} \right), \left(\bar{t} \rightarrow W^- + \bar{b}, W^- \rightarrow l_2^- \nu_{l_2} \right)  \, ,
\end{equation}
to calculate this normalisation factor since it provides the largest contribution to the SM background.  We generate events for this process at LO with {\tt MadGraph5\_aMC@NLO}, using {\tt NNPDF4.0}~\cite{NNPDF:2021njg} in the 4-flavour scheme.
As Fig.\ \ref{fig:direct_ATLAS} shows, the number of events is suppressed for larger values of $m_{T2}$, leading to poorer statistics and higher uncertainties in these bins.
Accordingly, we neglect higher bins and define the normalisation factor as the ratio of the ATLAS simulation to our LO {\tt MadGraph} simulation in the first bin only, which gives $K_1 = 0.478 \pm 0.076$. 
The uncertainty in $K_1$ is largely dominated by the uncertainty in the ATLAS SM background simulation. Therefore, generating more events for our LO background simulation would only lower the statistical uncertainty in our simulation but not the total uncertainty on the normalisation factor.

Fig.\ \ref{fig:direct_ATLAS} shows the curves for ALP signal + SM background for different values of $c_t$ compared to the data, assuming $f_a = 1$ TeV.  We choose values of $c_t$ around 1 because these are the values of $c_t$ to which the analysis is sensitive: the background + signal already overestimates the measured events, so larger couplings would lead to even more events. The blue filled area corresponds to the experimental uncertainty. As argued before, the {\tt Madgraph} uncertainty on the signal as well as the uncertainty on the background simulation is negligible with respect to the experimental uncertainty, so for better visibility we show only the latter uncertainty. 

We calculate the resulting constraints on $\left|c_t\right|$ as follows.
The ALP-$t$-$\bar{t}$ vertex is proportional to $c_t/f_a$, so the cross-section and therefore number of events scales with $(c_t/f_a)^2$.
We assume a Poisson likelihood, which can be expressed as
\begin{equation}
 \mathcal{L}(c_t) = \prod_{k=1}^{N_{\rm bins}} \frac{\exp \left( -\left( \left(\frac{c_t}{f_a}\right)^2s_k + b_k) \right) \right)\left(\left(\frac{c_t}{f_a}\right)^2 s_k+b_k\right)^{n_k}}{n_k!} \, ,
\end{equation}
where $s_k$, $b_k$ and $n_k$ denote the number of signal (for $c_t =1$, $f_a = 1$ TeV), background and data events, respectively.
$L(c_t)$ is maximised for no ALP signal at all, $c_t = 0$. We make use of the profile likelihood ratio to obtain a limit of 
\begin{equation}
\label{eq:dirlim}
    \left|\frac{f_a}{c_t}\right| > 552.2\, {\rm GeV} \qquad {\rm at}\,\, 95\% \,{\rm C.L.}
\end{equation}

In order to obtain combined limits on $c_t$, we also investigated a reinterpretation of a CMS search for top squarks in the $t \bar{t}$ + MET final state~\cite{CMS:2021eha}. Unfortunately, with the given phase space cuts we were not able to accurately generate data in the signal region, as further explained in Appendix~\ref{sec:direct_constraints_CMS}. We thus present only the limit from the ATLAS analysis.

\section{Indirect couplings to other SM particles}
\label{sec:indirect}
In this section we investigate two further constraints on the ALP-top coupling, which will end up being weaker than the direct constraint presented in Sec.~\ref{sec:direct}, but still significant. In Sec.~\ref{sec:ttbar} we will investigate the indirect limits that stem from the 
ALP-top contribution to the loop-induced gluon-gluon fusion to an ALP leading to top pair production. In particular we will derive two limits from the ATLAS and CMS analyses of high $p_T$ top pair production. In Sec.~\ref{sec:vvloop} we recast the loop-induced limits on the 
ALP to vector boson couplings into limits on $|c_t/f_a|$ according to the expressions given in Table~\ref{tab:recast}. 

\subsection{ALP-mediated $t \bar{t}$ production}
\label{sec:ttbar}
\begin{figure}
     \centering
     \begin{subfigure}[b]{0.49\textwidth}
         \centering
         \includegraphics[width=0.8\textwidth]{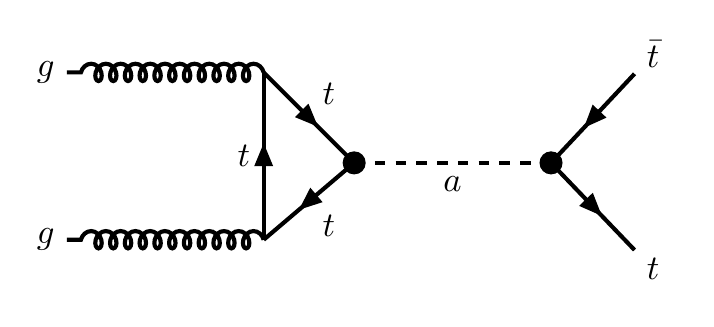}
         \caption{}
         \label{fig:ttbar_feynman}
     \end{subfigure}
     \begin{subfigure}[b]{0.49\textwidth}
         \centering
         \includegraphics[width=\textwidth]{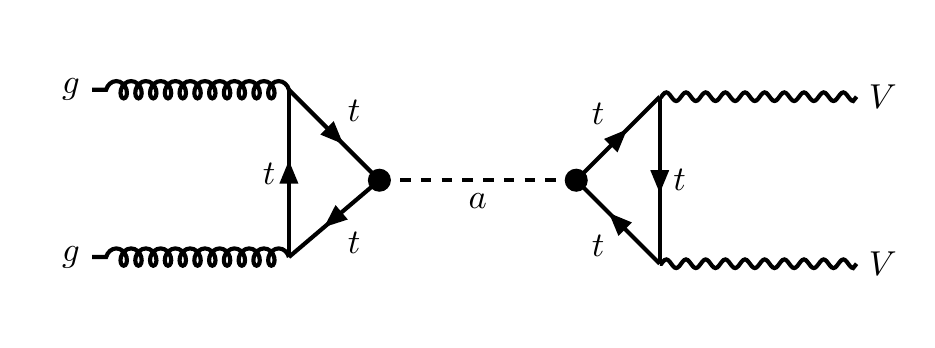}
         \caption{}
         \label{fig:diboson_feynman}
     \end{subfigure}
\caption{Contribution from the off-shell ALP to $ p p \rightarrow t \bar{t}$ production (left) and $ p p \rightarrow VV$ production (right), where $V$ denotes $\gamma$, $W^{\pm}$, $Z$, $g$.  The coupling $c_{t}$ to tops is indicated by the black circle.}
\label{fig:indirect_feynman}
\end{figure}

The ALP-top coupling $c_t$ may be indirectly probed through its contribution 
to loop-induced gluon-gluon fusion to an ALP leading to top quark pairs, as shown in Fig.~\ref{fig:ttbar_feynman}.
As discussed in Sec.~\ref{sec:theory}, $c_t$ induces an effective ALP-gluon-gluon coupling given by
\begin{equation}
    \mathcal{L}_{agg} =  c_{agg}^{\rm eff} \mathcal{O}_{\tilde{G}} = -\frac{a}{f_a} c_{agg}^{\rm eff}G_{\mu \nu}^{a} \tilde{G}^{a \mu \nu}  \, ,
\end{equation}
where $c_{agg}^{\rm eff}$ is related to $c_t$ by $c_{agg}^{\rm eff} = -\frac{\alpha_s}{8 \pi} c_t$, see Table.~\ref{tab:recast}.
Here we consider a light off-shell ALP contributing non-resonantly to the top quark pair production process\footnote{If the ALP is sufficiently heavy, this will instead lead to resonant ALP production, as discussed in Ref.~\cite{Bonilla:2021ufe}.}.  

We will estimate the impact of Fig.~\ref{fig:ttbar_feynman} by a tree-level calculation, approximating the effect of the top loop using the effective ALP-gg coupling $c_{agg}^{\rm eff} = -\frac{\alpha_s}{8 \pi} c_t$.
The presence of derivative operators in $\mathcal{O}_{\tilde{G}}$ will lead to additional factors of momentum in the amplitude and cross section numerator, enhancing the $\hat{s}$-dependence of the ALP-mediated $t \bar{t}$ cross section, relative to the SM.
We confirm this by calculating the partonic cross section for the ALP contribution to $t \bar{t}$ production, finding that it scales with $\hat{s}$ as
\begin{equation}
    \hat{\sigma}_{\rm ALP}(\hat{s}) \sim \frac{c_t^2 c_{\tilde{G}}^2 m_t^2}{f_a^4} \left(1 - \frac{2 m_t^2}{\hat{s}}\right) \, .
\end{equation}
Conversely, the SM cross section decays with $\hat{s}$.  
As a result, we can expect the ALP to lead to an enhancement in the $m_{t \bar{t}}$ distribution tails, 
as noted in Ref.~\cite{Gavela:2019cmq,Carra:2021ycg} for ALP-mediated diboson production.

However, the ALP-SM interference cannot be neglected.  In fact, the contribution from the interference term $\sigma_{\rm SM-ALP}$ will only be negligible relative to the ALP cross section $\sigma_{\rm ALP-ALP}$
 for very large values of $c_t$, due to the factor of $\alpha_s / 8 \pi$ in the effective ALP-$gg$ vertex.  Furthermore, we find that $\sigma_{\rm SM-ALP}$ decays with $m_{t \bar{t}}$ more quickly than the SM.  This is a result of the fact that the ALP diagram only interferes with the SM $t-$ and $u$-channel diagrams; the SM $s$-channel diagram contains a colour factor $f^{abc}$ which vanishes when contracted with the colourless ALP coupling.
Therefore, while the partonic interference cross section
scales as
\begin{equation}
    \hat{\sigma}_{\rm SM-ALP} (\hat{s}) \sim \frac{1}{\hat{s}} \log\left(\sqrt{\frac{\hat{s}}{m_t^2}} \right) \, ,
\end{equation}
for large-$\hat{s}$, the SM cross section receives an additional
contribution from the $s$-channel diagram, which decreases the rate of the cross section decay.
The result is that, although the $\hat{s}$ dependence of $\hat{\sigma}_{\rm SM-SM}$ and $\hat{\sigma}_{\rm ALP-SM}$ are proportional to each other at very high $\hat{s}$, the SM cross section decays more slowly with $\hat{s}$ than the interference term for values of $\hat{s}$ around 1-2 TeV, where the $s$-channel diagrams cannot be completely neglected.  

We refer to Appendix~\ref{app:ttbar-calc} for more details.  In particular, Fig.~\ref{fig:APP-ALPSMkinematics} shows the dependence on $m_{t \bar{t}}$ of the ALP and SM-ALP interference, and in Fig.~\ref{fig:APP-ALPSMsignal} we display the
impact of the ALP signal on the $m_{t \bar{t}}$ and $p_T$ spectra.  Note that the SM-ALP interference can be suppressed by considering high-$p_{T}$ top quarks, such as those measured by ATLAS in their analysis of top quark pair production at high transverse momentum~\cite{ATLAS:2022xfj}.  We will constrain the ALP-top coupling from this dataset in the what follows.

\begin{figure}[htb!]
     \centering
         \includegraphics[width=0.85\textwidth]{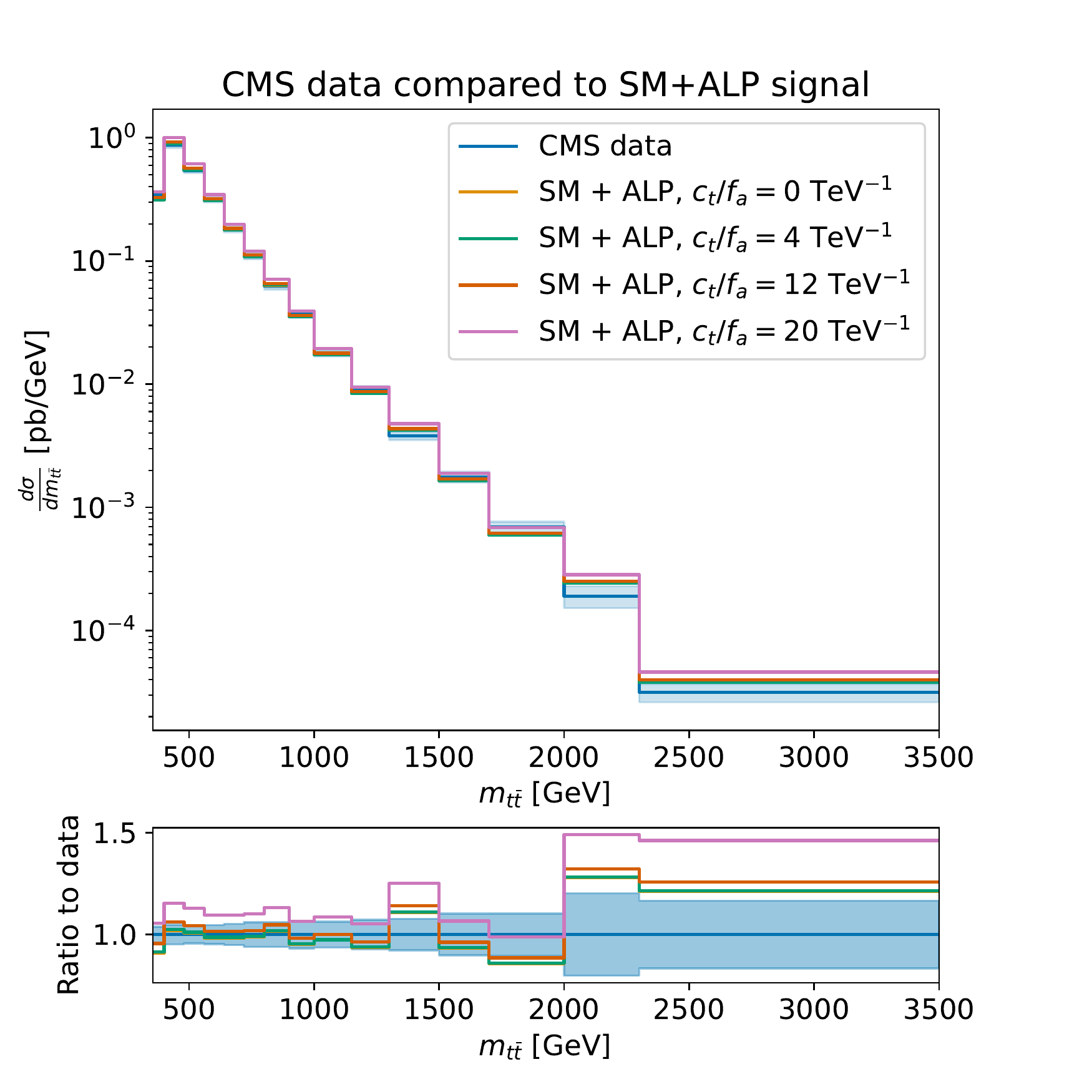}
        \caption{ALP signals ($f_a = 1$ TeV) compared to the CMS $m_{t \bar{t}}$ distribution~\cite{CMS:2021vhb}.}
\label{fig:CMS_tt_data}
\end{figure}

\paragraph{CMS measurement of top quark pair production.} 
We first study the signal produced by an off-shell ALP contribution to $t \bar{t}$ production and its impact on the measurement of $t \bar{t}$ production performed by CMS in Ref.~\cite{CMS:2021vhb}. Subsequently we will look at the impact on the measurement of $t \bar{t}$ production presented by ATLAS in Ref.~\cite{ATLAS:2022xfj}.  CMS provides a measurement of the $m_{t \bar{t}}$ distribution in the lepton+jets channel using 137 fb$^{-1}$ of LHC Run II data. The $m_{t \bar{t}}$ bins extend up to 3.5 TeV, and we make use of the data unfolded to the parton level and extrapolated to the full phase space, as shown in Fig.~\ref{fig:CMS_tt_data}.

The ALP signal is calculated as follows.  We first calculate the SM $t \bar{t}$ process at LO in {\tt MadGraph5\_aMC@NLO}~\cite{Alwall:2014hca} using {\tt NNPDF4.0}~\cite{NNPDF:2021njg} in the 5 flavour scheme.
Predictions at NNLO in QCD have been made available by several collaborations~\cite{Bernreuther:2004jv,Czakon:2015owf,Czakon:2020qbd,Catani:2020tko,Mazzitelli:2021mmm}. Here we calculate them by using the {\tt HighTea} public tool~\cite{Czakon:2023hls}.
We calculate a k-factor by comparison of the NNLO and LO predictions.
The ALP signal is then calculated at LO in QCD using {\tt MadGraph5\_aMC@NLO} and making use of the UFO model {\tt ALP\_linear\_UFO}~\cite{ALPUFO}.  The NNLO/LO QCD k-factor is then applied to the signal to account for the increase in normalisation due to the inclusion of missing NNLO QCD corrections.  Further details of the ALP signal and UFO model are discussed in Appendix~\ref{app:ALPtheory}.

Recall that the ALP-$gg$ coupling is related to $c_t$ by $c_{\rm agg}^{eff} = - \frac{\alpha_s}{8 \pi} c_t$. 
 Given a value of $c_t$, we evaluate $c_{agg}^{\rm eff}$ from $c_t$ while taking into account the scale dependence of $\alpha_s$.  In particular, we make use of \texttt{RunDec}~\cite{Herren:2017osy} to evaluate $\alpha_s$ at the scale of each $m_{t \bar{t}}^c$ bin, where $m_{t \bar{t}}^c$ denotes the centre of each bin in the $m_{t \bar{t}}$ measurement.

We determine the constraints on $c_t$ at 95\% CL by making use of a Gaussian likelihood, defined as
\begin{equation}
    \label{eq:chi2def}
    \mathcal{L}(c_t) \propto \textrm{exp} \big( -\frac{1}{2} \sum_{i,j=1}^{n_{\rm dat}} (T_i(c_t) - D_i) (V^{-1})_{ij} (T_j(c_t) - D_j) \big) \, ,
\end{equation}
where $D_i$ and $T_i$ denote the data points and their corresponding theory predictions respectively, and $V$ denotes the experimental covariance matrix including both statistical and correlated systematic uncertainties.

A comparison between the ALP signal and CMS data is shown in Fig.~\ref{fig:CMS_tt_data}.  We note that the ALP signal shows no energy-growing behaviour in the $m_{t \bar{t}}$ distribution tail, as previously motivated and further discussed in Appendix~\ref{app:ttbar-calc}. 
Even at large values of $|c_t/f_a|$ such as $|c_t/f_a| \sim 20$ TeV$^{-1}$ shown in Fig.~\ref{fig:CMS_tt_data}, the ALP signal leads to a small deviation in the $m_{t \bar{t}}$ tail.  However, the contribution at low $m_{t \bar{t}}$ is large as a result of the large interference term, as discussed previously and as it can be seen explicitly in Fig.~\ref{fig:APP-ALPSMkinematics}.  

In fact, the constraints we obtain from this data on $c_t$ are largely determined by the low-$m_{t \bar{t}}$ bins.  
A constraint of 
\begin{equation}
\label{eq:indcms}
  \left|\frac{f_a}{c_t}\right| \, > \,103.1 \,{\rm GeV} \qquad {\rm at}\,\,95\% {\rm C.L.}   
\end{equation}
 is obtained from this CMS $m_{t \bar{t}}$ measurement.
We investigate the dependence of this constraint on the high-$m_{t \bar{t}}$ bins by removing the bins above 800 GeV from our analysis, and find that the constraint on $c_t$ is relatively stable, increasing only slightly to $|c_t/f_a| > 9.8 \textrm{ TeV}^{-1}$.  We conclude that at the values of $m_{t \bar{t}}$ probed in this measurement, the ALP-SM interference effect dominates and we do not gain sensitivity to the ALP-top coupling from the $m_{t \bar{t}}$ distribution tails.

\begin{figure}[htb!]
\centering
 \includegraphics[width=0.8\textwidth]{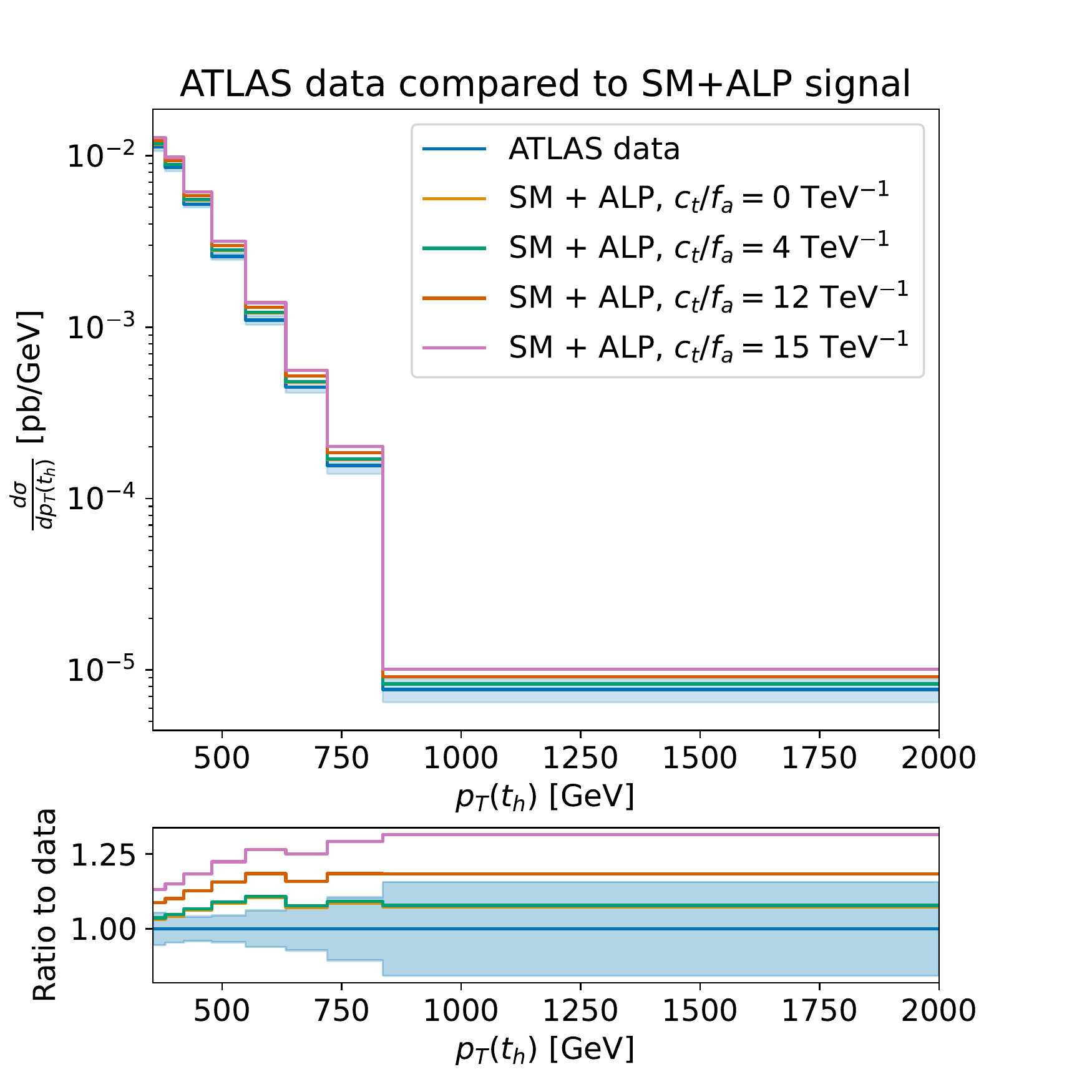}
\caption{ALP signals ($f_a = 1$ TeV) compared to the ATLAS boosted-top $p_{T}(t)$ spectrum~\cite{ATLAS:2022xfj}.}
\label{fig:ATLAS_tt_data_test}
\end{figure}

\paragraph{ATLAS measurement of top quark pair production with a high-$p_T$ top quark.} Next, we consider the impact of the ALP signal on an ATLAS measurement of $t \bar{t}$ production, this time considering a measurement in the lepton+jets channel with high-$p_T$ top quarks and $\mathcal{L}=$139 fb$^{-1}$~\cite{ATLAS:2022xfj}.  We consider the $p_T$ spectrum of the boosted hadronically decaying top quark, as shown in Fig.~\ref{fig:ATLAS_tt_data_test}.  In this case the data is provided in the fiducial region rather than the full phase space, and at particle level in the lepton + jets decay channel.  We estimate the effect of the ALP signal on the $p_{T}$ distribution as follows.
As in the previous analysis, we calculate the ALP signal at LO in QCD using {\tt MadGraph5\_aMC@NLO} and the UFO model {\tt ALP\_linear\_UFO}~\cite{ALPUFO}, applying a cut of $\eta_{t} <2$ and $p_{T}^t > 355$ GeV to approximate the fiducial region of the measurement.  We then apply a normalisation factor to account for the missing higher order QCD corrections.  This factor is found by taking the ratio of the NNLO QCD predictions, obtained directly from the ATLAS analysis, to the LO SM calculation obtained using {\tt MadGraph5\_aMC@NLO} in the presence of the phase space cuts above.

As in the previous case, we make use of a Gaussian likelihood, defined in Eq.~\eqref{eq:chi2def}, to obtain constraints on the ALP-top coupling $c_t$.
We find that $c_t$ is constrained by the $p_T$ spectrum to 
\begin{equation}
\label{eq:indatlas}
  \left|\frac{f_a}{c_t}\right| > 169.5 \,{\rm GeV} \qquad {\rm at}\,\,95\% {\rm C.L.}   
\end{equation}
Here we benefit from the use of top quarks boosted to $p_{T} > 355$ GeV, as further discussed in Appendix~\ref{app:ttbar-calc}.  In this kinematic regime the ALP cross section becomes non-negligible relative to the ALP-SM interference, and we find that the energy-growing effect of the ALP signal in the tail of the $p_{T}$ distribution provides the dominant source of constraints.  The result is a constraint on $c_t$ which is stronger than that resulting from the CMS $m_{t \bar{t}}$ distribution.  However, we note that in both cases we obtain constraints that are weaker than those found from the direct search for ALP-top couplings in $t \bar{t} +$ MET, as discussed in Sec.~\ref{sec:direct}.

\subsection{ALP-mediated diboson production via top couplings}
\label{sec:vvloop}
In Ref.~\cite{Gavela:2019cmq}, non-resonant searches in which an ALP is an off-shell mediator of a 2 → 2
scattering process were put forward for the first time.  They were proposed as a means to constrain the coupling of
the ALP with vector bosons through the contribution of the ALP to diboson production $g g \rightarrow VV$ by taking advantage of the large-$\hat{s}$ enhancement $\hat{\sigma} \sim \hat{s}/f_a^4$ of the ALP signal.  As shown in Fig.~\ref{fig:diboson_feynman}, the existence of a nonzero ALP-top coupling $c_t$ induces loop contributions to diboson production.  

In this section we recast the constraints obtained on the ALP-vector boson  couplings in Refs.~\cite{Gavela:2019cmq}, ~\cite{Carra:2021ycg} and~\cite{CMS:2021xor} onto constraints on $c_t$. We remind the reader that we work in the scenario in which $c_t$ is the only tree-level coupling of the ALP to the SM fields, and we estimate the loop diagram shown in Fig.~\ref{fig:diboson_feynman} by the tree-level amplitude with effective couplings given by those in Table~\ref{tab:recast}. 

In particular, we recast the bounds on $g_{agg}$, $g_{aZZ}$ and $g_{a\gamma\gamma}$ which were extracted from CMS searches for non-resonant particles at $\sqrt{s}=13$ TeV. The 95\% C.L. upper bounds derived from the searches in the $ZZ$ and $\gamma\gamma$ channels, valid up to $m_a\le 200$ GeV in Ref.~\cite{Gavela:2019cmq}, are given by
\begin{equation}
\label{eq:gazz}
    |g_{agg}\,g_{aZZ}| < 1\,{\rm TeV}^{-2}
\end{equation}
 and
 \begin{equation}
 \label{eq:gagg}
    |g_{agg}\,g_{a\gamma\gamma}| < 0.08\,{\rm TeV}^{-2}
\end{equation}
respectively\footnote{These numbers have been digitised from the plots in Fig.~4 of Ref.~\cite{Gavela:2019cmq}}. Using the expressions given in Table~\ref{tab:recast} and the conversion factor of Eq.~\eqref{eq:ctog}, and evaluating $\alpha_S(Q)$ at the scale of $Q\sim 1.5$ TeV (which is the  $p_T$ region in the vector-boson pair production that most strongly constrains the $g_{aVV'}$ couplings), these translate into 
\begin{align}
     & \left|\frac{f_a}{c_t}\right| > 3.5 \,{\rm GeV} \qquad {\rm at}\,\,95\% {\rm C.L.}  \label{eq:recast_gazz}\\
         & \left|\frac{f_a}{c_t}\right| > 22.5 \,{\rm GeV} \qquad {\rm at}\,\,95\% {\rm C.L.}  \label{eq:recast_gagg} 
\end{align}
respectively. 
In the same work, using a CMS search in the di-jet angular distribution, and including both the ALP signal and its interference with the SM, a limit on $g_{agg}$ was derived.  However, this limit is outside the regime of validity of the EFT and will not be discussed here. 

In Ref.~\cite{Carra:2021ycg}, the analysis of~\cite{Gavela:2019cmq} is complemented by computing the constraints on $g_{aWW}$ and $g_{aZ\gamma}$ couplings  that can be derived from ATLAS non-resonant searches in the $WW$ and $Z\gamma$ final stated. The 95\% C.L. exclusion limits, valid up to $m_a\le 100$ GeV, are
 \begin{align}
 \label{eq:gaww}
   |g_{agg}\, g_{aWW}| & < 0.62\,\,{\rm TeV}^{-2}\\
   \label{eq:gazg}
   |g_{agg}\, g_{aZ\gamma}| &< 0.37\,\,{\rm TeV}^{-2}.
\end{align}
 Given that according to our choice of the ALP-top coupling the $WW$ contribution vanish, we only recast Eq.~\eqref{eq:gazg} into a limit on $c_t$ and obtain
 \begin{align}
     & \left|\frac{f_a}{c_t}\right| > 11.0 \,{\rm GeV} \qquad {\rm at}\,\,95\% {\rm C.L.}  \label{eq:recast_gazg}
\end{align}

In Ref.~\cite{CMS:2021xor}, further constraints on the coupling $g_{aZZ}$ are obtained from a CMS non-resonant search in a $ZZ$ final state.  The constraint at 95\% CL, valid for $m_a \le 3$ TeV, is given by
\begin{align}
 \label{eq:gazzCMS}
   |g_{agg}\, g_{aZZ}| & < 0.64\,\,{\rm TeV}^{-2},
\end{align}
which translates to the following constraint on $|f_a/c_t|$:
\begin{align}
     & \left|\frac{f_a}{c_t}\right| > 17 \,{\rm GeV} \qquad {\rm at}\,\,95\% {\rm C.L.}  \label{eq:recast_gazz_new}
\end{align}
Finally we observe that in Ref.~\cite{Bonilla:2022pxu} non-resonant ALP-mediated vector-boson scattering is considered, where the ALP participates as an off-shell mediator. However with current data the bounds are weaker than those extracted from VV' non-resonant searches. Hence, given that the bounds obtained in this section are already much weaker than those obtained in Sec.~3 and in Sec.~4.1, we will not include those.
\subsection{Lower-energy precision measurements}

Various precise SM measurements would be modified due to the presence of a light state, the ALP, coupled to the SM through the top, see Ref.~\cite{Aiko:2023trb} for a thorough discussion on how the ALP would contribute to precision observables. In our case, with an ALP with direct couplings only to the top, the limits weaken considerably. For example, 
contributions to the anomalous magnetic moment $g-2$ would be proportional to loop-induced $g_{a\gamma\gamma}$ and loop-induced and mass suppressed $g_{a\ell^+\ell^-}^2$~\cite{Bonilla:2022qgm}.

In this section we discuss those measurements that could compete with the collider probes we have already considered. In particular we will find that ALP-induced FCNC are very sensitive probes of the ALP coupling, albeit only in a small low-mass region.  

\subsubsection{Limits from Flavour-Changing Neutral Currents induced by the ALP}
\label{sec:flavor}
Precise limits on the rare Kaon and B-meson decays can be used to set bounds on the ALP. In particular, for an invisible axion, the relevant searches are transitions from $K\to \pi$+invisible and $B\to K$+invisible.

The latest measurement of $K \to \pi \nu \bar \nu$~\cite{NA62:2021zjw} from NA62 is setting limits on new $X$ particles contributing to the $K$ decay of 
\begin{equation} \label{Kdec1}
    BR(K\to \pi + X) \lesssim (3-6) \times 10^{-11} \textrm{ at 90\% CL for } m_a < 110 \textrm{ MeV} 
\end{equation}
and
\begin{equation}\label{Kdec2}
    BR(K\to \pi + X) \lesssim   10^{-11} \textrm{ at 90\% CL for } m_a \in [160,260]  \textrm{ MeV} 
\end{equation}
and for the B-decays~\cite{BaBar:2013npw} the current best limit still comes from BaBar,
\begin{equation}\label{Bdec}
    BR(B\to K + \textrm{ inv.}) < 3.2 \times 10^{-5} \textrm{ at 90\% CL for } m_a \lesssim 5  \textrm{ GeV} 
\end{equation}
although Belle II already reached competitive values of 4.1$\times 10^{-5}$~\cite{Kurz:2022rsg} and should reach values of ${\cal O}(10^{-6})$~\cite{Belle-II:2022cgf} with 1 ab$^{-1}$.

\begin{figure}[h!]
    \centering
    \includegraphics{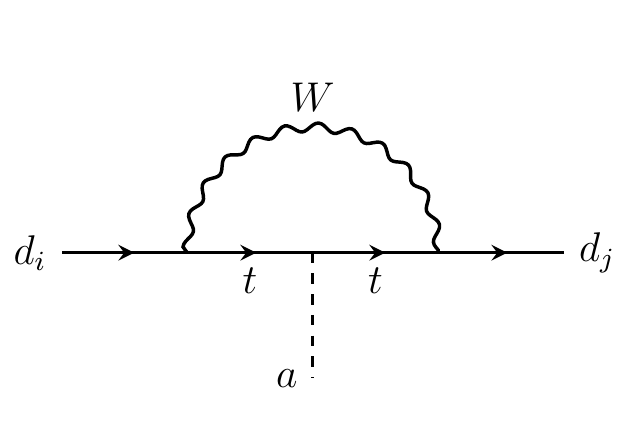}
    \caption{Loop diagram mediating  rare decays to ALPs through the $t\bar t$  coupling $c_t$.}
    \label{fig:FCNC}
\end{figure}
These decays would be mediated by a loop with a top and a W as virtual states, and where the top would radiate an ALP, see Fig.~\ref{fig:FCNC}. This diagram would be proportional to $|V_{ts} \, V_{tq}|^2 \, \left(\frac{c_t}{f_a}\right)^2$, where $q=d$ or $b$ for the Kaon and B-meson decays, respectively.

The explicit expressions for the  widths can be obtained by adapting the discussion in Ref.~\cite{Izaguirre:2016dfi} to the case of the top-ALP coupling. They would be given by 
\begin{eqnarray}\label{eq:alpFCNC}
    \Gamma(K^+ \to \pi^+ \, a) & = & \frac{m_{K^+}^3}{64 \pi} \, \left(1-\frac{m_{\pi^+}^2}{m_{K^+}^2}\right)^2 \, |g_{asd}|^2 \, \lambda_{\pi^+ a}^{1/2} \\
     \Gamma(B \to K \, a) & = & \frac{m_{B}^3}{64 \pi} \, \left(1-\frac{m_{K}^2}{m_{B}^2}\right)^2 \, |g_{abs}|^2 \, f_0^2(m_a^2) \,\lambda_{K a}^{1/2}
\end{eqnarray}
where 
\begin{equation}
    g_{as(d,b)} \simeq 1.2 \, \frac{c_t}{f_a}\, \frac{3 \sqrt{2} G_F m_W^2}{16 \pi^2} \, V_{ts} \, V^*_{t(d,b)}.
\end{equation}
The lifetime of  the $K^+$ and $B$ can be obtained from ~\cite{ParticleDataGroup:2004fcd}, and the value of the scalar form factor can be obtained from Ref.~\cite{Ball:2004rg}.

Other contributions, where the ALP would be radiated by a W would be enhanced by CKM factors with respect to the diagram with the coupling $t\bar t  a$, but would be negligible as  the $W^+ W^- a$ coupling  is zero, or suppressed by an additional loop factor, as explained in Sec.~\ref{sec:theory}.

From Eqs.~\ref{eq:alpFCNC}, we can compare with the curent NA62~\ref{Kdec1} ~\ref{Kdec2} and Babar limits~\ref{Bdec} to obtain mass-dependent limits on $c_t/f_a$ from Kaon decays
\begin{eqnarray}
    \left|\frac{c_t}{f_a}\right| \lesssim 2.8\times 10^{-4} \textrm{ GeV}^{-1} \textrm{ for }m_a < 110 \textrm{ MeV and }  m_a \in [160,260]  \textrm{ MeV, } 
\end{eqnarray}
and from B-meson decays
\begin{eqnarray}
    \left|\frac{c_t}{f_a}\right| \lesssim 1.5\times 10^{-6} \textrm{ GeV}^{-1} \textrm{ for }  m_a \lesssim 5  \textrm{ GeV.} 
\end{eqnarray}

\subsubsection{Electroweak Precision tests}

The ALP would contribute to the $Z$-width through a decay~\cite{Aiko:2023trb}
\begin{equation}
   Z\to a + \gamma  \ . 
\end{equation}
The $Z$ width is know down to a precision of $\simeq 2\times 10^{-3}$ GeV~\cite{ParticleDataGroup:2020ssz}.

In our case, the ALP contribution would be suppressed by an electroweak loop factor, 
\begin{equation}
    \Gamma(Z\to a \gamma)=\frac{m_Z^3 \alpha_{em}^2}{216 \pi^3}\,  \left(1-\frac{m_a^2}{m_Z^2}\right)^3
\end{equation}
and would lead to a very weak limit of the order of $f_a/c_t \lesssim 2$ GeV.

Similarly, the Higgs could decay to an ALP and a photon or a Z boson through a top triangle, proportional to a loop factor and $(y_t \alpha_{em})^2$. As the total width of the Higgs is not measured at the same level as the $Z$, the corresponding limit on the top coupling is even weaker. 

The same conclusion can be drawn from other precision measurements. Despite their precise determination, these electroweak parameters 
are not sensitive to the scenario that we consider in this paper, when we compare with the LHC probes.  

\section{EFT validity}
\label{sec:validity}

So far, we have obtained limits on the scale and coupling of the ALP to tops using LHC probes. These are high-energy handles on the ALP, in a kinematic regime where one should ask the question of whether the use of the  Effective Field Theory description is adequate. 

To evaluate the validity range for this description, we have to understand whether the limit we have placed in the combination of ALP parameters corresponds to a region where the scale of the EFT expansion, $f_a$, is larger than the typical $p^2$ of the signature we used; namely we have to ask the question: 
\begin{equation}
   \textrm{is the limit on } \left|\frac{f_a}{c_t}\right| \textrm{  consistent with } f_a > \sqrt{\hat s} \textrm{ ? }\ ,
\end{equation}
where $\hat s$ denotes the typical $p^2$ of the events we used to set the limit.

\begin{figure}[t!]
\centering
\includegraphics[width=0.48
\textwidth]
{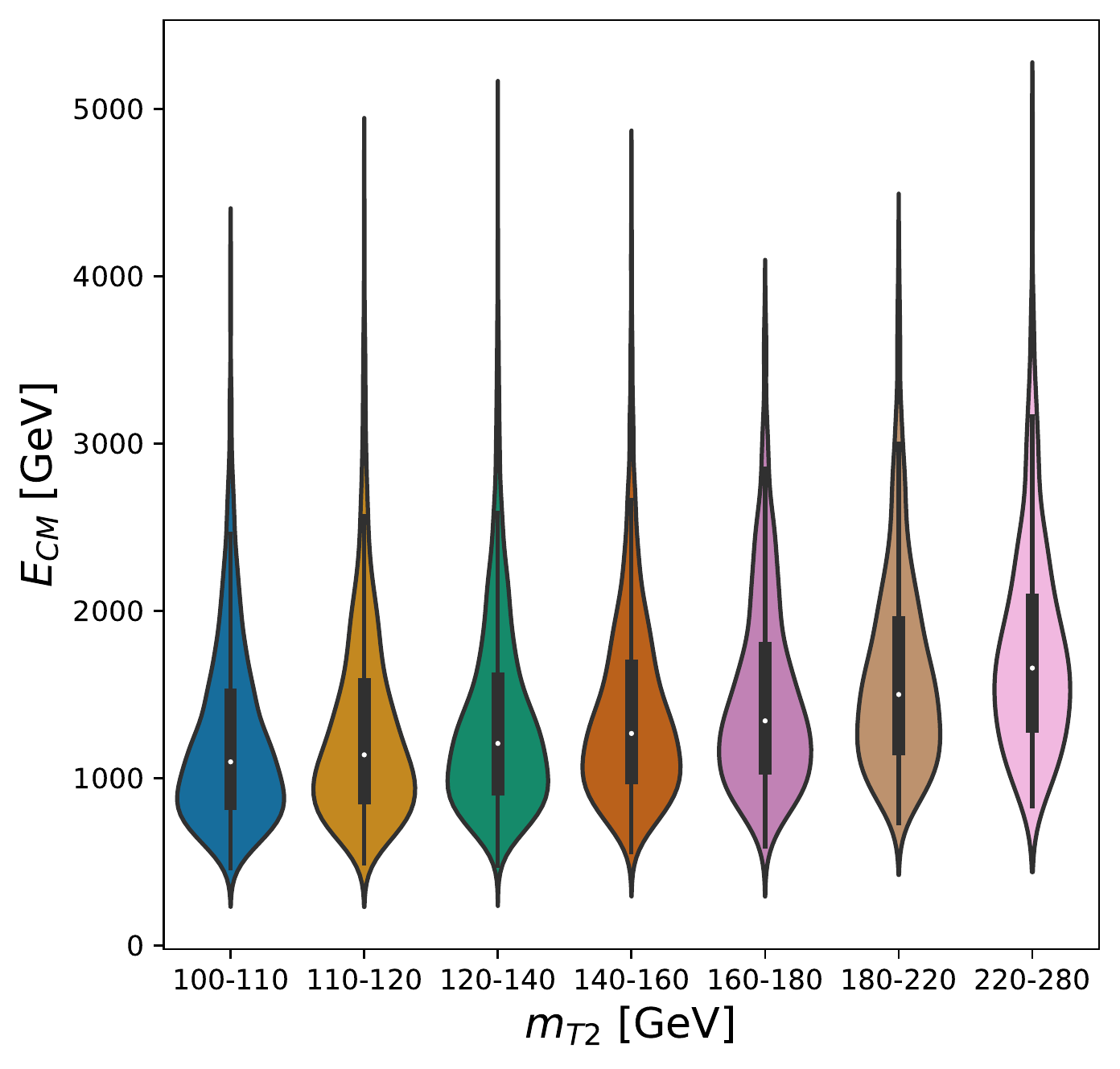}
\includegraphics[width=0.48\textwidth]
{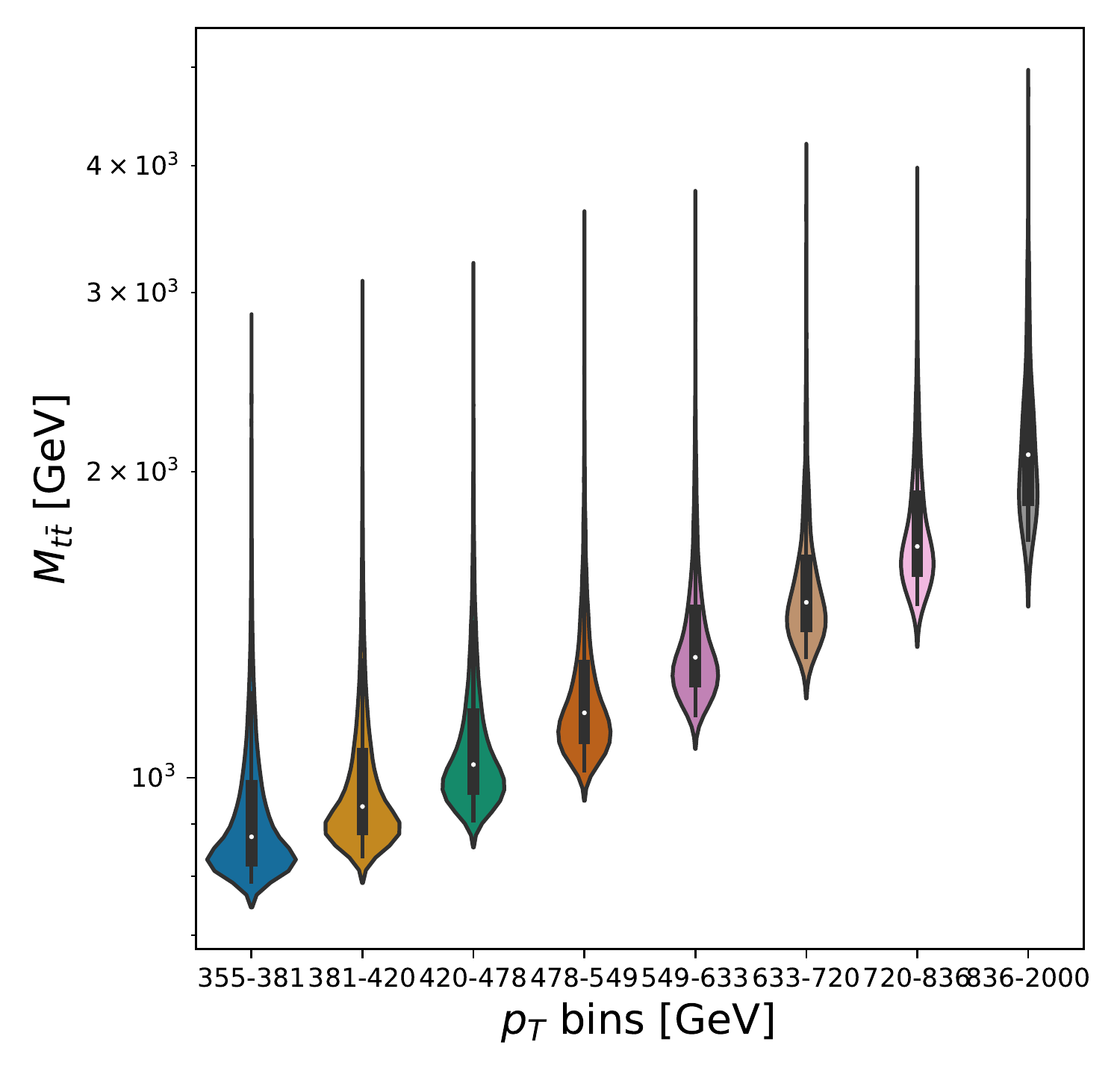}
\caption{\textbf{Left panel}: Distribution of $E_{CM}$ in each $m_{T2}$ bin of the ATLAS measurement for $t\bar{t}$ + MET (direct search). \textbf{Right panel:} Distribution of $m_{t\bar{t}}$ in each $p_T$ bin for the ATLAS search with boosted top quarks. In both cases the mean and standard deviation of $E_{CM}$ in each bin are represented by the white dot and black bar respectively and the coloured distributions show the probability density of $E_{CM}$ events in each bin, taken directly from the {\tt MadGraph5\_aMC@NLO} simulation, cf. Sec.~\ref{sec:direct_constraints_ATLAS} and~\ref{sec:ttbar}.}
\label{fig:atlas_axion_correlation_mt2}
\end{figure}

To illustrate this issue, we first turn our attention to the leading channel in our analysis,  where we have direct access to the ALP-top coupling: $t\bar t$+MET. To suppress backgrounds, this channel is defined with strict kinematic cuts on the final leptons and b-jets. In particular, a hard cut on $m_{T_2}(\ell\ell)>$ 110 GeV is applied. This and other selection cuts shape the kinematic range $\hat s$ of the events.

In Fig.~\ref{fig:atlas_axion_correlation_mt2} we illustrate the distribution of $E_{CM} = \sqrt{\hat{s}}$ for each $m_{T2}$ bin in the ATLAS search. We observe a slight correlation towards higher values of $E_{CM}$ for higher $m_{T2}$ bins, but as the constraints are dominated by the first three bins we take the mean value $E_{CM} = 1000$ GeV as an estimate of the relevant scale $\sqrt{\hat{s}}$.
Therefore, we should interpret the limit on $|f_a/c_t|$ (GeV) as a limit where $f_a$ would have to be above 1 TeV. As one can see from Fig.~\ref{fig:summary}, that would imply that the analysis would be valid for a value of $c_t\gtrsim 2$. 

Alternatively, one could address the issue of validity by  truncating the analysis at different values of $m_T$, as discussed in Ref.~\cite{Mimasu:2014nea}.

Note, though, that these comments on validity are based on a  reinterpretation of a search tailored to a different physical situation, SUSY, and that with a dedicated analysis and higher luminosity one should expect to be sensitive to smaller couplings $c_t$.

Next, we consider the validity of the EFT in the context of the indirect constraints on $c_t$ obtained from the non-resonant production of an ALP in $t \bar{t}$ production, as discussed in Sect.~\ref{sec:ttbar}.  In Eq.~\ref{eq:indcms} a constraint of $|f_a/c_t|>103.1$ GeV was obtained from a CMS measurement of the $t \bar{t}$ invariant mass spectrum, in which the highest bin is given by $m_{t \bar{t}} \in [2.3,3.5]$ TeV.  Taking the midpoint of this bin, $m_{t \bar{t}} = 2.9$ TeV, as an estimate of $\sqrt{\hat{s}}$, we find that this constraint on $|f_a/c_t|$ is valid for values of $c_t\gtrsim28$.  

Similarly, a constraint of $|f_a/c_t|>169.5$ GeV was obtained from the analysis of an ATLAS measurement of the $p_T$ spectrum of boosted top quarks, as given in Eq.~\ref{eq:indatlas}.  To obtain an estimate of the relevant scale $\sqrt{\hat{s}}$, we display in Fig.~\ref{fig:atlas_axion_correlation_mt2} the relationship between $m_{t \bar{t}}$ and the $p_T$ of each bin.  In particular, for each bin in $p_T$ we determine the distribution of $m_{t \bar{t}}$ from a sample of 100000 simulated events.  As expected, we observe a strong correlation between $p_T$ and $m_{t \bar{t}}$, with a mean value of $m_{t \bar{t}} = 2$ TeV in the final bin.  
We take this value as an estimate of the scale $\sqrt{\hat{s}}$.  As a result, the constraint in Eq.~\ref{eq:indatlas} is valid for values of $c_t\gtrsim12$.

\section{Other ALP signatures sensitive to the top coupling: contact interactions and long-lived particles}
\label{sec:others}

In this paper we have mainly focused on collider-stable ALPs, an assumption that depends on the ALP mass and the typical boost in our experimental signatures, see Appendix~\ref{sec:stability} for a discussion on the ALP collider stability.

In this section we would like to briefly discuss other experimental signatures, not based on the collider stability assumption. For example, the limits obtained in Sec.~\ref{sec:indirect} from searches for non-resonant ALP production do not rely on a collider-stable ALP, therefore the limit we obtained in this paper would carry forward in the case of a decaying ALP.

A decaying ALP or an off-shell ALP could contribute to channels with clean final states, for example leading to diphoton or dilepton signatures which are, in principle, very powerful handles on new physics. 

Unfortunately, the limits on ALP couplings coming from dilepton final states would be negligible due to the mass suppression of the ALP coupling to fermions. For example, in the non-resonant case,  analyses such as Ref.~\cite{CMS:2018nlk} searching for contact four-fermion quark-lepton interactions, would apply. These limits are expressed as effective operators of the type $\frac{4 \, \pi}{\Lambda^2} (\bar q \gamma^\mu q) (\bar \ell \gamma_\mu \ell)$, with different choices of chiralities. The current bounds are  of the order of $\Lambda\sim 30$ TeV.   
The four-fermion coupling due to an exchange of an off-shell ALP would be very suppressed, roughly scaling as $m_q m_\ell c_q c_\ell /\hat s f_a^2$.  Moreover, both the coupling to light quarks $c_q$ and to leptons, electrons or muons, $c_\ell$ would be further loop-suppressed in the context of this paper.  

Searches for diboson resonances would be more powerful than dileptons. For example, bump searches in diphoton final states would be sensitive to ALPs through their coupling to gluons and photons via top loops. As an example, one can reinterpret a recent ATLAS search for low-mass (10-70 GeV) ALPs in  Ref.~\cite{ATLAS:2022abz}.   Nevertheless, in the context of this paper, the loop suppression of these couplings would render current searches less sensitive than our $t\bar t+a$ channel.

Finally, we would like to mention the natural possibility that the ALP leads to Long-Lived Particle (LLP) signatures. Those final states are  increasingly interesting signatures for a mature experiment like the LHC, see Ref.~\cite{Alimena:2019zri} for the LLP community white paper, which explores the huge range of new possibilities for LLPs at colliders and their interplay with other probes. 

LLPs are typical of pseudo-Goldstone bosons, such as ALPs or gravitinos, because these states couple through higher-order suppressed interactions, which would typically lead to long lifetimes. In our case, and even in the region of the parameter space where the ALP could be collider-stable and the main results of this paper would apply, there could be an interesting interplay between the missing energy signatures and LLP searches. The reason is that the ALP boost from the $t\bar t\,  a$ final state is not fixed, and from the same production mechanism there would  be a kinematic region where the ALP would decay inside the detector. Therefore, a combined analysis of $t\bar t$+MET signatures with LLP associated to tops would have an enhanced sensitivity to the ALP. A detailed study of the interplay between missing energy and LLP signatures is left for future work.

\section{Conclusions}\label{sec:concls}

In this paper we studied the current sensitivity of the LHC to a light axion-like particle (ALP) coupled to top pairs. As the ALP coupling to fermions is proportional to the fermion mass, there is a strong motivation to focus on third generation fermions at the LHC, and in particular on the top quark. 

We found an interesting interplay between ALP-strahlung signatures in which the ALP is radiated by one of the top quarks, and off-shell production of the ALP into a $t\bar t$ final state. 
We conclude that if the ALP escapes detection, the $t\bar t$+MET signature is currently more sensitive than the off-shell production, even in the high-$m_{t\bar t}$ channel. Nevertheless, one should keep in mind that the scaling with luminosity of these two experimental handles could be different, and in the future the high-$m_{t\bar t}$ could become more sensitive than the SUSY-like $t\bar t$+MET. 

\begin{figure}[tb]
    \centering
    \includegraphics[width=0.99\textwidth]{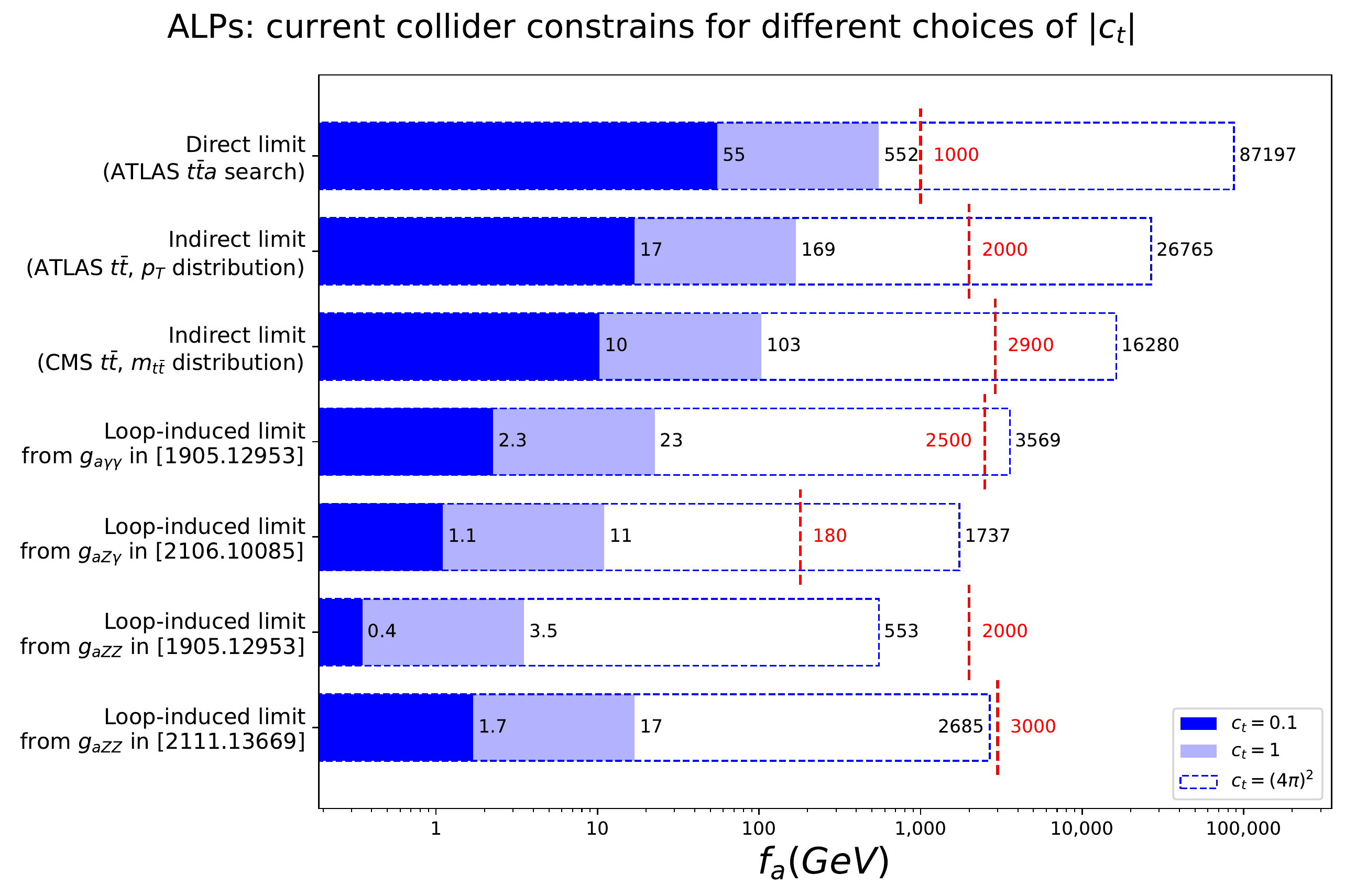}
    \caption{Summary plot of the most significant constraints on ALP-top  couplings presented in this work.  The bar for the direct limit corresponds to Eq.~\eqref{eq:dirlim}. The bars for the indirect constraints correspond to the limits stemming from the contribution to loop-induced gluon-gluon fusion to an ALP leading to $t\bar{t}$ production derived from the ATLAS and CMS $t\bar{t}$ measurements, given in Eqs.\eqref{eq:indcms}-\eqref{eq:indatlas}. Finally, the four lowest bars correspond to the recasting of the loop-induced $g_{aVV'}$ limits, given in Eqs.~\eqref{eq:recast_gazz}, \eqref{eq:recast_gagg}, \eqref{eq:recast_gazg} and \eqref{eq:recast_gazz_new}. The dark blue, light blue and dashed bars show the lower limits on $f_a$ for weak couplings $c_t = 0.1$, natural couplings $c_t =1$ and strong couplings $c_t = (4 \pi)^2$ respectively. The red dashed lines denotes the relevant energy scale $\sqrt{\hat{s}}$ for the respective process. As discussed in Sec.~\ref{sec:validity} $f_a > \sqrt{\hat{s}}$ is required for a valid EFT description.}
    \label{fig:summary}
\end{figure}
\begin{figure}[tb]
    \centering
    \includegraphics[width=0.99\textwidth]{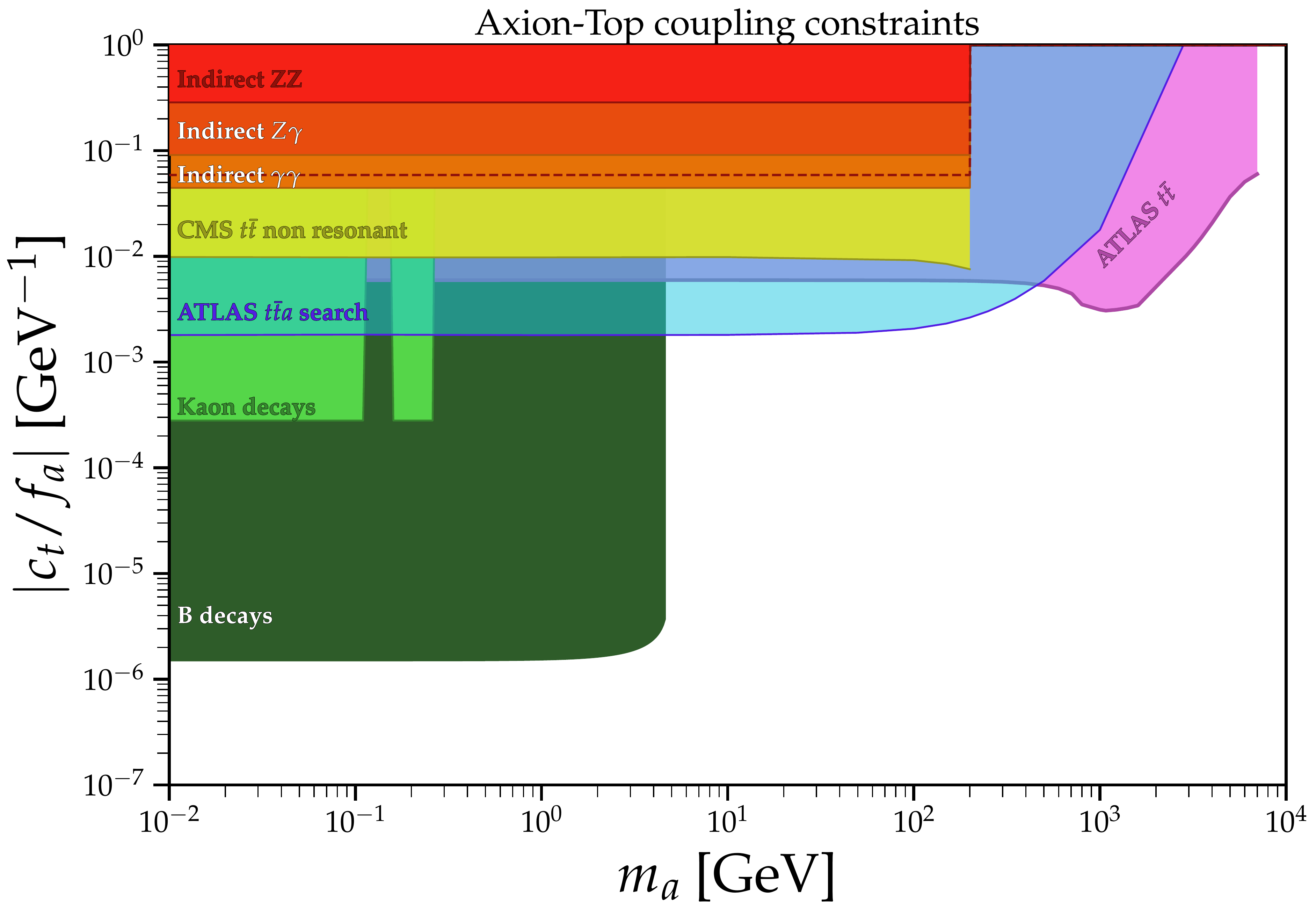}
    \caption{Summary plot of the constraints on ALP-top couplings presented in this work. The red, orange and yellow regions in the top-left corner correspond to the recasting of the loop-induced $g_{aVV'}$ limits, given in Eqs.~\eqref{eq:recast_gazz}, \eqref{eq:recast_gagg}, \eqref{eq:recast_gazg}, with the dashed red line corresponding to the most recent constraint from $g_{aZZ}$ of Eq.~\eqref{eq:recast_gazz_new}. The light blue area corresponds to the direct limits discussed in Sec.~\ref{sec:direct}. The light and dark green regions correspond to the flavour constraints 
    discussed in Sec.~\ref{sec:flavor}. The yellow region stems from the contribution to loop-induced gluon-gluon fusion to an ALP leading to $t\bar{t}$ production derived from the CMS $t\bar{t}$ measurements in the non-resonant region (below $200$~GeV), while the pink region is derived from the ATLAS $t\bar{t}$ measurements in the non-resonant region and in the resonant region.}
    \label{fig:summary2D}
\end{figure}

In our analysis we performed a re-interpretation of the SUSY search for stops ($t\bar t$+MET) and of the SM measurements of $t\bar t$ production at high invariant mass, as well as a recasting of the limits on the ALPs to vector boson couplings, see Figs.~\ref{fig:summary} and~\ref{fig:summary2D} for the summary of our results. 
In Fig.~\ref{fig:summary} we compare the limits we obtained from these different approaches. Note that all limits are given in terms of $|f_a/c_t|$, so in order to evaluate the EFT validity $f_a > \sqrt{\hat{s}}$ we present the lower limits on $f_a$ for three choices of $c_t$: weak ($c_t = 0.1$), natural ($c_t =1$) and strong ($c_t = (4 \pi)^2$) couplings. In Sec.~\ref{sec:validity} we motivated how to obtain the relevant energy scales $\sqrt{\hat{s}}$ for the individual processes which are shown as red dashed lines.  We find that even for the direct search, given currently the most stringent limits, we still need strong couplings of order $c_t =2$ to fulfill $f_a > \sqrt{\hat{s}}$. However, we expect that with higher luminosity the constraints will be more precise and push the regime of valid EFT descriptions into the range of natural couplings.  

In Fig.~\ref{fig:summary2D} we plot the limits that we identify in this analysis in the $(m_a,|c_t/f_a|)$ plane. We observe that up to $m_a=m_b\approx 4.7$ GeV the constraints coming from $B-$decays measurements by the BaBar collaboration~\cite{BaBar:2013npw} dominate over the other constraints that we considered in this paper. 
On the other hand, for values of $m_a$ above $m_b$, and up to $\approx 200$ GeV, the re-interpretation of the SUSY search for stops ($t\bar t$+MET) in terms of 
a direct search for $t\bar{t}a$ gives the strongest constraint, followed by the CMS and ATLAS indirect constraints from $t\bar{t}$ measurements in the non-resonant 
region and the recasting of the loop-induced $g_{aVV'}$ limits discussed in Sec.~\ref{sec:indirect}. When we reach the $t\bar{t}$ resonant region, then the ATLAS measurement 
of top quark pair production with a high-$p_T$ top quark gives the dominant constraint.
Note that in the re-interpretation of the SUSY search for stops ($t\bar t$+MET) in terms of 
a direct search for $t\bar{t}a$ we assumed that the ALP is collider-stable. According to the considerations of Sec.~\ref{sec:stability} 
this assumption holds only up to $m_a \lesssim 200$ MeV, so the light blue region corresponding to the $t\bar{t}a$ direct search should be seen as an interpolation 
above $200$ MeV. However, the collider stability assumption could be restored by allowing for a branching ratio of the ALP to an invisible sector.

In the paper we also discussed other signatures which could be added to this analysis, especially the long-lived particles (LLPs) signatures of an ALP. 
Although our study shows that the current LHC sensitivity is moderately good, $|f_a/c_t| \sim {\cal O}$(500 GeV), a dedicated experimental analysis, focusing on the particularities of the ALP signature, should lead to a more reliable and stronger bound, and could be carried over towards the high-luminosity LHC phase. For example, if the ALP effective theory could be incorporated alongside the SUSY interpretation of the $t\bar t$+MET final state, the issue of the validity could be properly studied there, including the adequate choice of the $m_{T_2}$ bins. Moreover, the ALP interpretation of the SM top measurements could be added in parallel to the Top SMEFT interpretation of these SM precision measurements, as in both cases the best sensitivity comes from the non-resonant boosted top region. Finally, one could consider the effect of including the ALP as part of the proton wave-function, as it was done in~\cite{McCullough:2022hzr} in the case of dark photon, and check what effects this would have on the ALP constraints.

\section*{Acknowledgements}

We would like to thank Ilaria Brivio for her help with the UFO implementation of the fermionic couplings, to Quico Botella and Miguel Nebot for their insights into FCNCs and to Jorge de Troconiz for pointing out a new CMS analysis. We would also like to acknowledge discussions with Rene Poncelet and Anke Biekötter.
FE is supported by the Generalitat Valenciana with the grant
GRISOLIAP/2020/145. The research of VS is supported by the Generalitat
Valenciana PROMETEO/2021/083 and the Ministerio de Ciencia e
Innovacion PID2020-113644GB-I00.
The work of M. M. and M. U is supported by the European Research
Council under the European Union’s Horizon 2020 research and innovation Programme (grant agreement
n.950246) and in part by STFC consolidated grant ST/T000694/1.
The work of M. U. is also funded by the Royal Society grant DH150088.  The research of M.~M. is also supported by the Deutsche Forschungsgemeinschaft (DFG, German Research Foundation) under grant 396021762 – TRR 257 Particle Physics Phenomenology
after the Higgs Discovery.

\appendix

\section{ALP Theory \& UFO model}
\label{app:ALPtheory}
Here we will discuss further theoretical background to our work.  We will introduce the ALP-top coupling in Sec.~\ref{subsec:theoryapp}, and outline how this coupling can be approximated in {\tt MadGraph} simulations from the \texttt{ALP\_linear\_UFO}~\cite{ALPUFO} model in Sec.~\ref{subsec:ufomodel}.

\subsection{Theoretical details}
\label{subsec:theoryapp}
A complete and non-redunant CP-even Lagrangian describing the interactions of the ALP with the SM up to $\mathcal{O}(1/f_a)$ is given as~\cite{Georgi:1986df,Bonilla:2021ufe,Brivio:2017ije}
\begin{align}
\label{eq:app-lagrangian}
    \mathcal{L} = \mathcal{L}_{\rm SM} + \mathcal{L}_{a}^{\rm total}
\end{align}
where $\mathcal{L}_{\rm SM}$ denotes the SM Lagrangian, and $\mathcal{L}_{a}^{\rm total}$ is given by
\begin{align}
\mathcal{L}_{a}^{\rm total} = \frac{1}{2} \partial_{\mu} a \partial^{\mu} a + \frac{1}{2} m_a^2 a^2 + c_{\tilde W} \mathcal{O}_{\tilde{W}}
+ c_{\tilde B} \mathcal{O}_{\tilde{B}}
+ c_{\tilde G} \mathcal{O}_{\tilde{G}}
+ \sum_{f = u,d,e} c_{f} \mathcal{O}_{f}
+ \sum_{f=Q,L} c_{f} \slashed{\mathcal{O}}_{f} \, .
\end{align}
The ALP couples anomalously to the SM gauge bosons:
\begin{align}
    \mathcal{O}_{\tilde{W}} &= -\frac{a}{f_a} W_{\mu \nu}^\alpha \tilde{W}^{\alpha \mu \nu}\\
    \mathcal{O}_{\tilde{B}} &= -\frac{a}{f_a} B_{\mu \nu} \tilde{B}^{\mu \nu}\\
    \mathcal{O}_{\tilde{G}} &= -\frac{a}{f_a} G_{\mu \nu}^a \tilde{G}^{a \mu \nu} \, ,
\end{align}
whereas its couplings to the SM fermions are described by the following shift-invariant interactions:
\begin{equation}
    \mathcal{O}_{f} = \frac{\partial_{\mu} a}{f_a} \bar{f} \gamma^{\mu} f \, ,
\end{equation}
where $f \in \{ Q_{L}, L_{L}, e, u, d \}$.  We denote by $\slashed{\mathcal{O}}_f$ the fact that the $i=j=1$ component of $\mathcal{O}_{Q}^{ij}$ and all three diagonal components of $\mathcal{O}_{L}$ have been removed to account for redundancies amongst the operators.  The coupling of the ALP to the Higgs, parametrised by the operator $\mathcal{O}_{a \Phi} = \frac{\partial^{\mu} a}{f_a} (\Phi^\dagger i \overset{\text{\scriptsize$\leftrightarrow$}}{D}_{\mu} \Phi)$ is not present in the Lagrangian of Eq.~\ref{eq:app-lagrangian}; this operator is redundant with the fermionic operators and has been traded for the ALP-fermion couplings using the equations of motion.  In total, there are 29 degrees of freedom.

The flavour-diagonal couplings of the ALP to the top quark are given by
\begin{align}
    \mathcal{L}_{a}^{\rm total}  &\supset \frac{\partial_{\mu} a}{f_a} \big( (c_{u})^{33}\bar{t}_{R} \gamma^{\mu} t_{R}  + (c_{Q})^{33} \bar{t}_{L} \gamma^{\mu} t_{L} \big) \\
    &\supset \frac{\partial_{\mu} a}{2 f_a}  (c_{u} - c_Q)^{33}\bar{t} \gamma^{\mu} \gamma^{5} t \, ,
\end{align}
where in the second line we have expressed the coupling in terms of $t = (t_L, t_R)^T$, and the vector coupling vanishes due to the pseudoscalar nature of the ALP.  We will denote the coefficient of this axial coupling by $c_t = (c_u - c_Q)^{33}$ as in Ref.~\cite{Bonilla:2021ufe} and write the Lagrangian of our simple top-specific scenario as
\begin{equation}
\label{eqn:topcouplingapp}
    \mathcal{L}_{\rm top} = c_t \frac{\partial_{\mu} a}{2 f_a} \bar{t} \gamma^{\mu} \gamma^5 t \, ,
\end{equation}
see also Eq.~\ref{eq:top_coupling}.

This ALP-top coupling $c_t$ is exactly the coupling that appears in the top-loop corrections to the ALP-bosonic couplings as calculated in Ref.~\cite{Bonilla:2021ufe}, for example
\begin{align}
    \hat{g}_{a \gamma \gamma}^{eff} &= -\frac{\alpha}{\pi} \frac{c_t}{f_a} Q_t^2 N_C B_1 \Big( \frac{4 m_t^2}{p^2} \Big)\\
    &=-\frac{4 \alpha}{3 \pi } \frac{c_t}{f_a} B_1 \Big( \frac{4 m_t^2}{p^2} \Big)\\
    &\rightarrow -\frac{4 \alpha}{3 \pi} \frac{c_t}{f_a} \hspace{10pt} \textrm{ for } p^2 \rightarrow \infty \, .
\end{align}

\subsection{Use of the bosonic ALP UFO model}
\label{subsec:ufomodel}
We will make use of the \texttt{ALP\_linear\_UFO} model~\cite{ALPUFO} to simulate the effect of this coupling on our \texttt{Madgraph} simulations of $g g \rightarrow a \rightarrow t \bar{t}$ and $p p \rightarrow t \bar{t} a$.  However, this UFO model is an implementation of the bosonic interactions of the ALP with the SM, neglecting fermionic interactions; in particular, it implements the following Lagrangian:
\begin{equation}
\label{eq:bosonicL}
\mathcal{L} = \mathcal{L}^{\textrm{SM}} + \frac{1}{2} (\partial_{\mu} a)(\partial^{\mu} a) + \frac{1}{2} m_a^2 a^2 + c_{\tilde{W}} \mathcal{O}_{\tilde{W}} + c_{\tilde{B}} \mathcal{O}_{\tilde{B}} +c_{\tilde{G}} \mathcal{O}_{\tilde{G}} +c_{a \Phi} O_{a \Phi} \, .
\end{equation}
We wish to take advantage of the relation between the operator $\mathcal{O}_{a \Phi}$ and the fermion-ALP couplings, approximating the effect of the coupling $c_t$ using $c_{a \Phi}$ in our simulations.  To do so, first consider a field redefinition of the Higgs doublet $\Phi$ and of the fermion fields given by
\begin{align}
    \Phi &\rightarrow \textrm{exp} (i x_{\Phi} a/f_{a}) \Phi\\
    Q_{L} &\rightarrow \textrm{exp}(i x_{Q_{L}} a/f_{a} ) Q_{L}\\
    t_{R} &\rightarrow \textrm{exp}(i x_{t_{R}} a/f_{a} ) t_{R}\\
    b_{R} &\rightarrow \textrm{exp}(i x_{b_{R}} a/f_{a} ) b_{R} \, ,
\end{align}
where $x_{\Phi} \in \mathbb{R}$ and each $x_{f} \in \mathbb{R}$.  This induces a shift in the SM Lagrangian given by~\cite{Bonilla:2021ufe,Brivio:2017ije} 
\begin{align}
\begin{split}
    \Delta \mathcal{L}_{\rm SM} &= -x_{\Phi} \mathcal{O}_{a \Phi} - \sum_{f \in Q_{L}, t_{R}, b_{R}} x_f \mathcal{O}_{f}
    + \frac{g_s^2}{32 \pi^2} \mathcal{O}_{\tilde{G}} \textrm{Tr} [2 x_Q - x_{t_{R}} - x_{b_{R}}]\\
    &+ \frac{g'^2}{32 \pi^2} \mathcal{O}_{\tilde{B}} \textrm{Tr}[\frac{1}{3} x_Q - \frac{8}{3} x_{t_{R}} - \frac{2}{3} x_{b_{R}}] + \frac{g^2}{32 \pi^2} \mathcal{O}_{\tilde{W}} \textrm{Tr}[3 x_Q]\\
    &+ \big[ (- x_{\Phi} Y_e) \mathcal{O}_{e \Phi} + (x_Q Y_d - Y_d x_{b_{R}} - x_{\Phi} Y_d) \mathcal{O}_{d \Phi} \\
    &+ (x_Q Y_u - Y_u x_{t_{R}} + x_{\Phi} Y_u) \mathcal{O}_{u \Phi}  + h.c. \big] \, .
\end{split}
\end{align}
First, we may choose $x_{\Phi} = c_{a \Phi}$ to remove the operator $\mathcal{O}_{a \Phi}$ from the full Lagrangian.  Secondly, we may remove $\mathcal{O}_{u \Phi}$ by setting $x_Q - x_{t_{R}} = -x_{\Phi} = -c_{a \Phi}$.  Thirdly, we will remove corrections to $\mathcal{O}_{\tilde{G}}$ by selecting $2 x_Q - x_{t_{R}} - x_{b_{R}} = 0$ and therefore $x_Q - x_{b_{R}} = c_{a \Phi}$.  This also has the effect of removing $\mathcal{O}_{d \Phi}$, since it implies that $x_{Q} - x_{b_{R}} - x_{\Phi} =0$.  Finally, consider the remaining top couplings,
\begin{align}
\begin{split}
    \Delta \mathcal{L}_{\rm SM} &\supset - x_{Q} \mathcal{O}_{Q} - x_{t_{R}} \mathcal{O}_{t_{R}}\\
    &\supset -\frac{\partial_{\mu} a}{2 f_a} (x_t - x_Q) \bar{t} \gamma^{\mu} \gamma^5 t - \frac{\partial_{\mu} a}{2 f_a} (x_t + x_Q) \bar{t} \gamma^{\mu} t \, ,
\end{split}
\end{align}
where on the second line we have ignored couplings to $b$ quarks.  By selecting $x_{t_{R}} =- x_{Q}$, we remove the vector-like coupling, retaining only the axial coupling that we would like to describe as in Eq.~\ref{eqn:topcouplingapp}.

Overall, this choice of field redefinition leads to the following Lagrangian:
\begin{align}
\begin{split}
    \mathcal{L} &= \mathcal{L}^{\textrm{SM}} + \frac{1}{2} (\partial_{\mu} a)(\partial^{\mu} a) + c_{\tilde{W}} \mathcal{O}_{\tilde{W}} + c_{\tilde{B}} \mathcal{O}_{\tilde{B}} +c_{\tilde{G}} \mathcal{O}_{\tilde{G}}
    \\
    & - \frac{3 g^2}{64 \pi^2} c_{a \Phi} \mathcal{O}_{\tilde{W}} - \frac{ g'^2}{64 \pi^2} c_{a \Phi} \mathcal{O}_{\tilde{B}} - c_{a \Phi} Y_e \mathcal{O}_{e \Phi} \\
    &- c_{a \Phi} \frac{\partial_{\mu} a}{2 f_{a}} \bar{t} \gamma^{\mu} \gamma^5 t + c_{a \Phi} \frac{\partial_{\mu} a}{2 f_{a}} \bar{b} \gamma^{\mu} \gamma^5 b + 2 c_{a \Phi} \frac{\partial_{\mu} a}{2 f_{a}} \bar{b} \gamma^{\mu}  b   \, .
\end{split}
\label{eqn:redef-lagrangian}
\end{align}
The bosonic operators in the first line may simply be removed by setting their coefficients to zero, $c_{\tilde{G}} = c_{\tilde{W}} = c_{\tilde{B}} = 0$.  The operators in the second line could be removed by further 
field redefinitions in the quark or lepton sector; however we ignore this for now as they are not relevant to the processes of interest at tree-level.  Finally, we note that switching on the coupling $c_{a \Phi}$ will switch on $c_t$, and that by comparing Eq.~\ref{eqn:topcouplingapp} with the first term in the last line in Eq.~\ref{eqn:redef-lagrangian}, we may identify $c_t$ with $-c_{a \Phi}$. 

Couplings to the bottom quarks will also be switched on as shown in the other two terms in the last line of Eq.~\ref{eqn:redef-lagrangian}. These couplings have no effect on simulations of $g g \rightarrow a \rightarrow t \bar{t}$ and the analysis of Sec.~\ref{sec:indirect}.  Conversely, they will contribute to $b \bar{b} \rightarrow t \bar{t} a$ and the analysis discussed in Sec.~\ref{sec:direct}.  However, these couplings constitute very small corrections to the $t \bar{t} +$ MET signature.  The $b \bar{b}$ channel of $t \bar{t}$ production is subdominant and PDF-suppressed relative to the gluon-gluon channel.  Furthermore, the fermion-ALP couplings are proportional to the fermion mass, and therefore these couplings will be $m_b$-suppressed relative to the ALP-top coupling.  In practice, however, we neglect the impact of this coupling by performing the analysis of Sec.~\ref{sec:direct} using PDFs in the 4-flavour scheme.

\section{Collider stability of the ALP} \label{sec:stability}
In the analysis of Sec.~\ref{sec:direct} we assumed the ALP to be collider stable, i.e having a lifetime that permits it to escape the detector without decaying. As a result, it will manifest as missing transverse energy in the measurement. 

The length an ALP travels before decaying scales with its lifetime, and accordingly with the inverse of its decay width $\Gamma(a)$, as
\begin{equation}
    d = \tau \beta c = \frac{\hbar} {\Gamma(a)} \frac{|\vec{p}_a|}{m_a}c \, ,
\label{eq:decay_length_general}
\end{equation}
where $|\vec{p}_a|$ denotes the momentum of the ALP $a$. 

We define the ALP to be  {\it collider stable} if this distance is larger than the typical region where the detection would happen. This, in turn, depends on the final state to which the ALP decays. For example, if the ALP would decay to hadrons, these would lead to the cleanest traces in the hadronic calorimeter and could escape detection if the typical decay length was longer than the position of the hadronic calorimenter from the interaction point.

The branching ratio to an ALP final state depends on its couplings and mass.
For the discussion on the mass, we will distinguish the two cases $m_a \sim 1$ MeV and $m_a > 1$ MeV, and study the different decay channels individually.

{\bf $m_a \sim 1$ MeV:}
For this mass, any decays into charged leptons (and anything heavier) are kinematically not accessible. 
The remaining 3 channels are $a \rightarrow \nu \bar{\nu} \nu \bar{\nu}$, which will not differ from a pure $\slashed{E}_T$ contribution, $a \rightarrow \gamma \gamma$ and $a \rightarrow \gamma \nu \bar{\nu}$. 
For $a \rightarrow \gamma \gamma$ we find $d = \frac{16 \pi \bar{h} c}{m_a^4}\frac{|\vec{p}_a|}{g_{a\gamma\gamma}^2}$ which, for $m_a = 1$ MeV and the experimental constraint $g_{a\gamma \gamma} < 10^{-5} \text{ GeV}^{-1}$, leads to a decay length of $d > 10^8 \text{ m} \times \left(\frac{|\vec{p}_a|}{\text{GeV}}\right)$. 

Similarly, for $a \rightarrow \gamma \nu \bar{\nu}$, we find $d > 3.3\cdot 10^{27} \text{ m} \times \left(\frac{|\vec{p}_a|}{\text{GeV}}\right)$.
This implies that for masses of $m_a \sim 1$ MeV, either decay channel will lead to the ALP travelling distances much larger than the typical size of a detector ($\sim 10$ m) before decaying.

{\bf $m_a > 1$ MeV:} 
To discuss the ALP collider stability in this mass range, we must look at the ALP decay width, taking into account heavier final states. 
Increasing the ALP mass will open new decay channels into pairs of fermions once the ALP mass is above twice the final state fermion mass. 

In our model only $g_{att}$ is non-zero at tree-level, so all ALP decays, except for $a \rightarrow t \bar{t}$, will be loop-induced. 
The coupling strengths at 1-loop level for ALPs into fermion pairs have been computed in Ref~\cite{Bonilla:2021ufe}.
Because only $c_t$ is nonzero at tree-level, the only contribution that survives in Eq.\ (4.37) of Ref~\cite{Bonilla:2021ufe} comes from $D_{mix}^{c_t}$, which defines the following contribution to the fermion-ALP coupling: 
\begin{equation}
    c_{f}^{eff} = \frac{\alpha_{em}}{2 \pi}c_t D_{mix}^{c_t} \, .
\end{equation}
The ALP is taken to be on-shell, i.e.\ $p^2 = m_a$, and we focus on the intermediate mass range  in which $m_f^2 \ll p^2 \ll (M_Z^2, M_W^2, M_H^2)$ where $m_f$ stands for the mass of the final state fermion. In this range one finds (Eq.\ (4.64) in \cite{Bonilla:2021ufe})
\begin{equation}
    D_{mix}^{c_t} = - \frac{3 T_{3,f} m_t^2}{2 s_w^2 M_W^2} \log\left(\frac{\Lambda^2}{m_t^2}\right) \, .
\end{equation}
\begin{figure}[t!]
    \centering
    \includegraphics[scale=0.22]{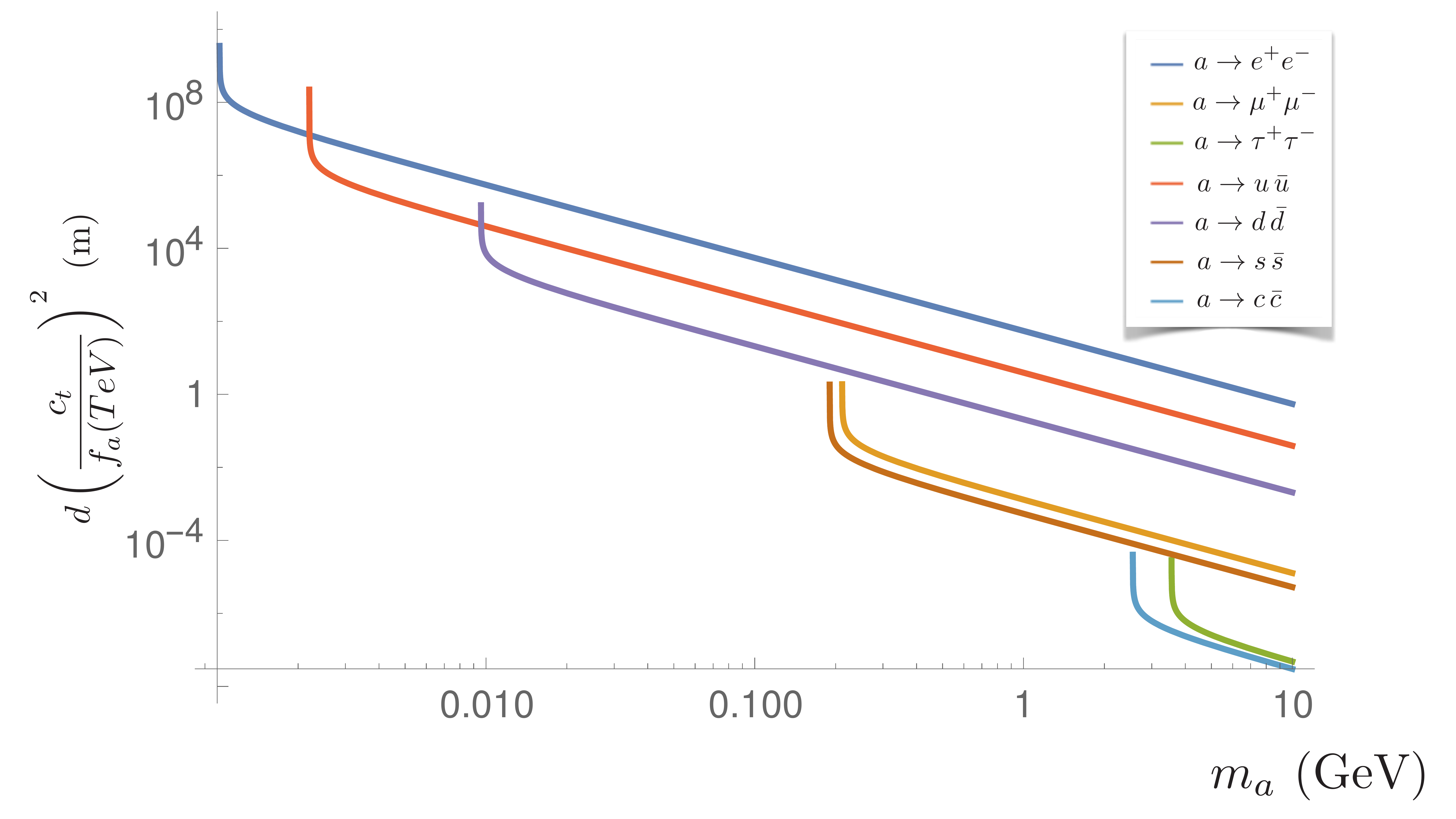}
    \caption{Decay lengths for ALPs into the $f\bar{f}$ pairs shown as a function of the ALP mass $m_a$ and assuming a boost factor $|\vec{p}_a|\simeq 100$ GeV.}
    \label{fig:decay_length}
\end{figure}
Further utilising
\begin{equation}
    \Gamma(a \rightarrow f \bar{f}) = \frac{N_C m_a m_f^2|c_{f}^{eff}|^2}{8 \pi f_a^2} \sqrt{1- \frac{4 m_f^2}{m_a^2}} \, ,
\end{equation}
we find the following:
\begin{equation}
    \Gamma(a \rightarrow f \bar{f}) = \frac{N_C m_a m_f^2}{8 \pi f_a^2} \sqrt{1- \frac{4 m_f^2}{m_a^2}}\left(\frac{\alpha_{em}}{2 \pi}\right)^2 |c_t|^2 \left(\frac{3 T_{3,f} m_t^2}{2 s_w^2 M_W^2} \log(m_t^2)\right)^2 \, .
\end{equation}
The decay length can then be calculated from Eq.~\ref{eq:decay_length_general}. Since the decay width for an ALP into a fermion pair dominates over the decay widths for $a \rightarrow \gamma \gamma$ and $a \rightarrow \gamma \gamma \nu \bar{\nu}$ in the considered mass range, we simply approximate $\Gamma(a) = \Gamma(a \rightarrow f\bar{f})$: 
\begin{equation}
d = \frac{\bar{h} c}{\Gamma(a \rightarrow f \bar{f}) m_a} \left(\frac{|\vec{p}_a|}{\text{GeV}} \right) \, .
\label{eq:decay_length_fermions}
\end{equation}

Note that for the missing energy signatures, the ALP momentum $|\vec{p_a}|$ is typically of the order of the minimal $\slashed{E}_T$ cut of $100 $ GeV. This motivates us to  set $|\vec{p_a}| = 100$ GeV to get a numerical approximation for the decay length in Eq.~\ref{eq:decay_length_fermions}. 

In Fig.~\ref{fig:decay_length} we plot the resulting decay length normalised by $c_t^2/f_a^2$ for a range of $f\bar{f}$ final states. For $f_a/c_t\sim {\cal O}$ (1 TeV), the decay length is in the range of the meter for ALPs with masses smaller than $\sim 200$ MeV. As the value of $f_a/c_t$ increases, so does the decay length, enlarging the region where the ALP would be collider stable. 

This plot is produced assuming that the typical ALP boost factor is given by $|\vec{p_a}| = 100$ GeV. ALPs produced with smaller or larger momenta could then lead to associated signatures with prompt or long-lived particle (LLP) signatures. These associated signatures could be additional handles on the ALP coupling to tops. For example, one could imagine combining the $t\bar t$+MET signature discussed in this paper with prompt $t\bar t+\gamma\gamma$ and LLPs in association with prompt tops. Such combined analysis is beyond the scope of this paper, but it would be a worthwhile study to increase the sensitivity to the ALP parameter space.

\section{Further details of the $\mathbf{t \bar{t}}$+MET analysis}
\label{sec:direct_constraints_CMS}
For the purpose of obtaining combined exclusion limits on $c_t$, we aimed to perform the same analysis presented in Sec.\ \ref{sec:direct}, reinterpreting a similar CMS search for top squarks~\cite{CMS:2021eha}.
CMS provides a measurement of the $p_T^{miss}$ distribution in the same 2 leptons + 2 jets + MET final state. The bins extend in equidistant $10$ GeV steps from $55$ GeV up to $295$ GeV. The signal region cuts are listed in Table \ref{tab:cms_cuts}.
\begin{table}[]
    \centering
    \begin{tabular}{c|c}
    \hline
    parameter & value \\
    \hline \hline
        $p_T$ leading lepton & $> 25$ GeV  \\
        $p_T$ subleading lepton & $> 20$ GeV \\ 
        $p_T^{miss}$ & $> 50$ GeV  \\
        \hline
        $m_{ll}$ & $ > 20$ GeV \\
        $m_{T2}(ll)$ & $> 80$ GeV  \\
        $|m_Z - m_{ll}|$ & $ > 15$ GeV  \\
        \hline
        $|\eta|$ leptons & $< 2.4$ \\
        $\Delta R = \sqrt{(\Delta \eta)^2 + (\Delta \Phi)^2}$ & $ = 0.3$ for electrons \\
        & $ = 0.4$ for muons \\  
        \hline
        $\cos( \Delta \Phi (p_T^{miss}, j))$ & $< 0.80$ for leading jet \\
        & $< 0.96$ for subleading jet \\
        \hline
    \end{tabular}
    \caption{Phase space cuts defining the signal region in the CMS search~\cite{CMS:2021eha}.}
    \label{tab:cms_cuts}
\end{table}
The azimuthal angle between the missing transverse momentum and one jet is defined as
\begin{equation}
    \cos(\Delta \Phi) =  \frac{\vec{p}_T^{miss}\cdot \vec{p}_T^{j}}{|\vec{p}_T^{miss}| |\vec{p}_T^{j}|} = \frac{p_x^{miss} p_x^j + p_y^{miss} p_y^j}{\sqrt{\left( (p_x^{miss})^2 + (p_y^{miss})^2 \right) \left( (p_x^{j})^2 + (p_y^{j})^2 \right)}}.
\end{equation}

However, generating $pp \rightarrow t\bar{t}$ SM background events with the CMS cuts listed in Table \ref{tab:cms_cuts} with  {\tt MadGraph5\_aMC@NLO}, we notice that the events drop off sharply at $m_{T2} \sim m_{W}$, where $m_{W}$ denotes the W boson mass.   Thus, cutting at the precise value of $m_{T2} = 80$ GeV removes a huge amount of events.  Moreover, the analysis is very sensitive to the cut, with variations around $m_{T2} = 80$ GeV leading to very different efficiencies.
Therefore we do not pursue the reinterpretation of this CMS dataset, and present instead the limits on $c_t/f_a$ obtained from the direct ATLAS search in Sec.~\ref{sec:direct}.

\FloatBarrier
\section{Non-resonant ALP contribution to $\mathbf{t \bar{t}}$ production}
\label{app:ttbar-calc}

\begin{figure}
     \centering
     \begin{subfigure}[b]{0.4\textwidth}
         \centering
         \includegraphics[width=\textwidth]{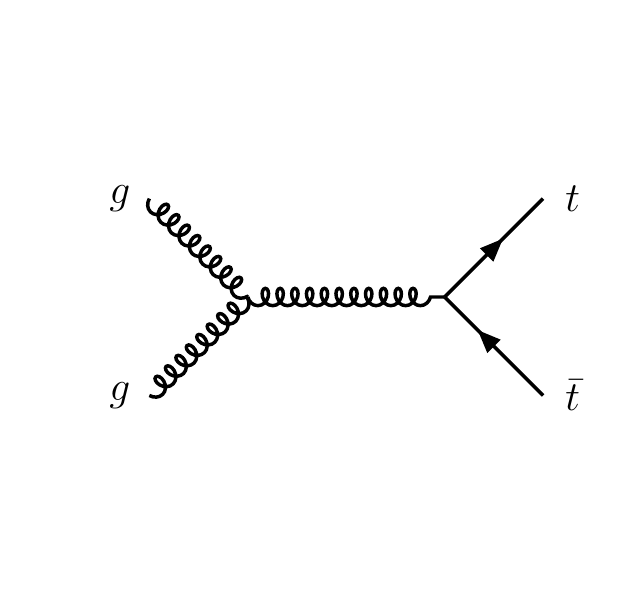}
         \caption{SM $s$-channel}
     \end{subfigure}
     \begin{subfigure}[b]{0.4\textwidth}
         \centering
         \includegraphics[width=\textwidth]{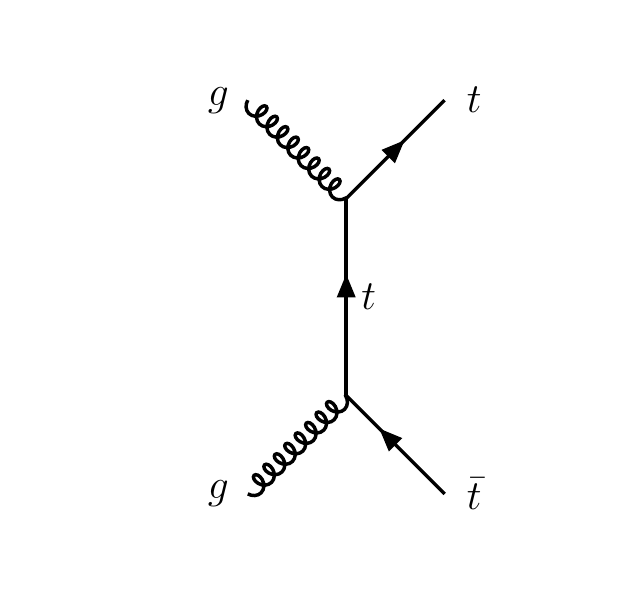}
         \caption{SM $t$-channel}
     \end{subfigure}
     \begin{subfigure}[b]{0.4\textwidth}
         \centering
         \includegraphics[width=\textwidth]{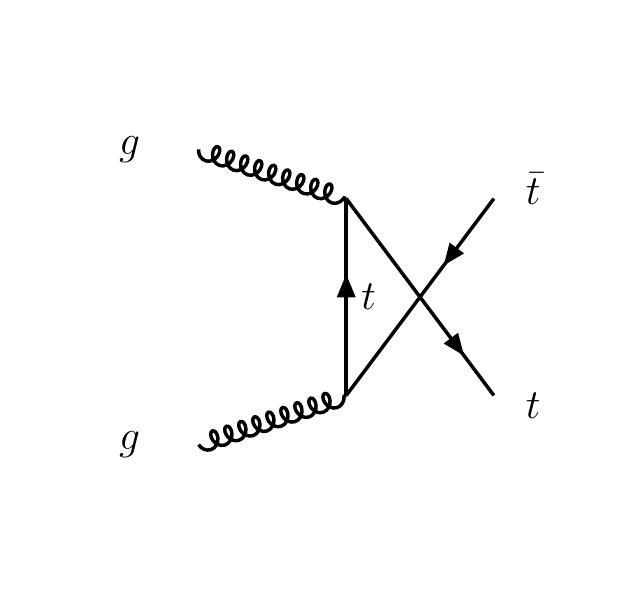}
         \caption{SM $u$-channel}
     \end{subfigure}
     \begin{subfigure}[b]{0.4\textwidth}
         \centering
         \includegraphics[width=\textwidth]{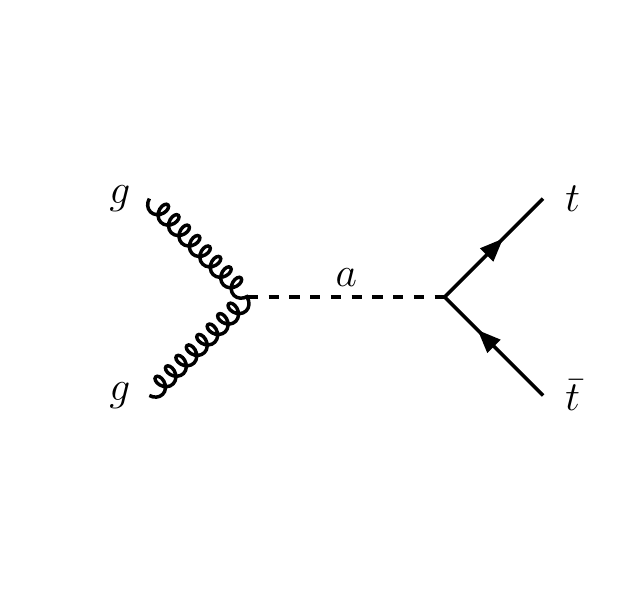}
         \caption{ALP contribution}
     \end{subfigure}
        \caption{Relevant diagrams for the calculation of the SM-ALP interference in $gg \rightarrow t\bar{t}$ production.}
        \label{fig:axion-sm-interference_diagrams}
\end{figure}

We present here details of the calculations of the ALP contribution to the process $g g \rightarrow t \bar{t}$ at LO in QCD.  The relevant Feynman diagrams that contribute to the interference between the SM process in the gluon-gluon channel and the ALP-mediated process are shown in Fig.~\ref{fig:axion-sm-interference_diagrams}.  We approximate the top-loop shown in Fig.~\ref{fig:ttbar_feynman} by an effective ALP-$gg$ coupling parametrised by $c_{agg} = -\frac{\alpha_s}{8 \pi} c_t$, and perform a tree-level calculation of the process $g g \rightarrow a \rightarrow t \bar{t}$. The Feynman rules for vertices involving ALPs are given in the Appendix of Ref.\ \cite{Brivio:2017ije}.  Omitting the overall colour and numerical factors, we obtain the following expressions for the spin-averaged matrix element squared of the ALP term and the SM-ALP interference:
\begin{align}
    |\mathcal{M}_{\rm ALP-ALP}|^2 &\propto \alpha_s^2 m_t^2 \frac{c_t^4}{ f_a^4} \frac{\hat{s}^3}{(\hat{s}-m_a^2)^2}\\
    |\mathcal{M}_{\rm SM-ALP}|^2 &\propto \alpha_s^2 m_t^2   \frac{ c_t^2}{f_a^2} \frac{\hat{s} (\hat{t} - \hat{u})^2}{(\hat{s} - m_a^2)(\hat{t} - m_t^2)(\hat{u} - m_t^2)} \, ,
\end{align}
where $\hat{s}, \hat{t}, \hat{u}$ are the Mandelstam variables defined as
\begin{equation}
\begin{split}
    s & = (p_1 + p_2)^2 = 2 (p_1 \cdot p_2) \\
    t & = (p_1 - p_3)^2 = -2 (p_1 \cdot p_3) + m_t^2 \\
    u & = (p_1 - p_4)^2 = - 2(p_1 \cdot p_4) + m_t^2,
\end{split}    
\end{equation}
where $p_1,p_2$ are the momenta of the incoming gluons and $p_3,p_4$ are the momenta of the outgoing top quark and antiquark. 
 Note the difference in denominators between the ALP-ALP and SM-ALP terms:  the ALP-ALP contribution to $t \bar{t}$ is a purely s-channel diagram; however it interferes with only the SM $t-$ and $u$-channels. This is because the ALP is colourless, leading to a factor of $\delta^{ab}$ in its coupling to a pair of gluons.  The SM $s$-channel proceeds through the totally antisymmetric triple gauge coupling $\propto f^{abc}$, and hence the interference between the ALP diagram and the SM $s$-channel is zero.

\begin{figure}[htb!]
     \centering
     \begin{subfigure}[b]{0.49\textwidth}
         \centering
         \includegraphics[width=\textwidth]{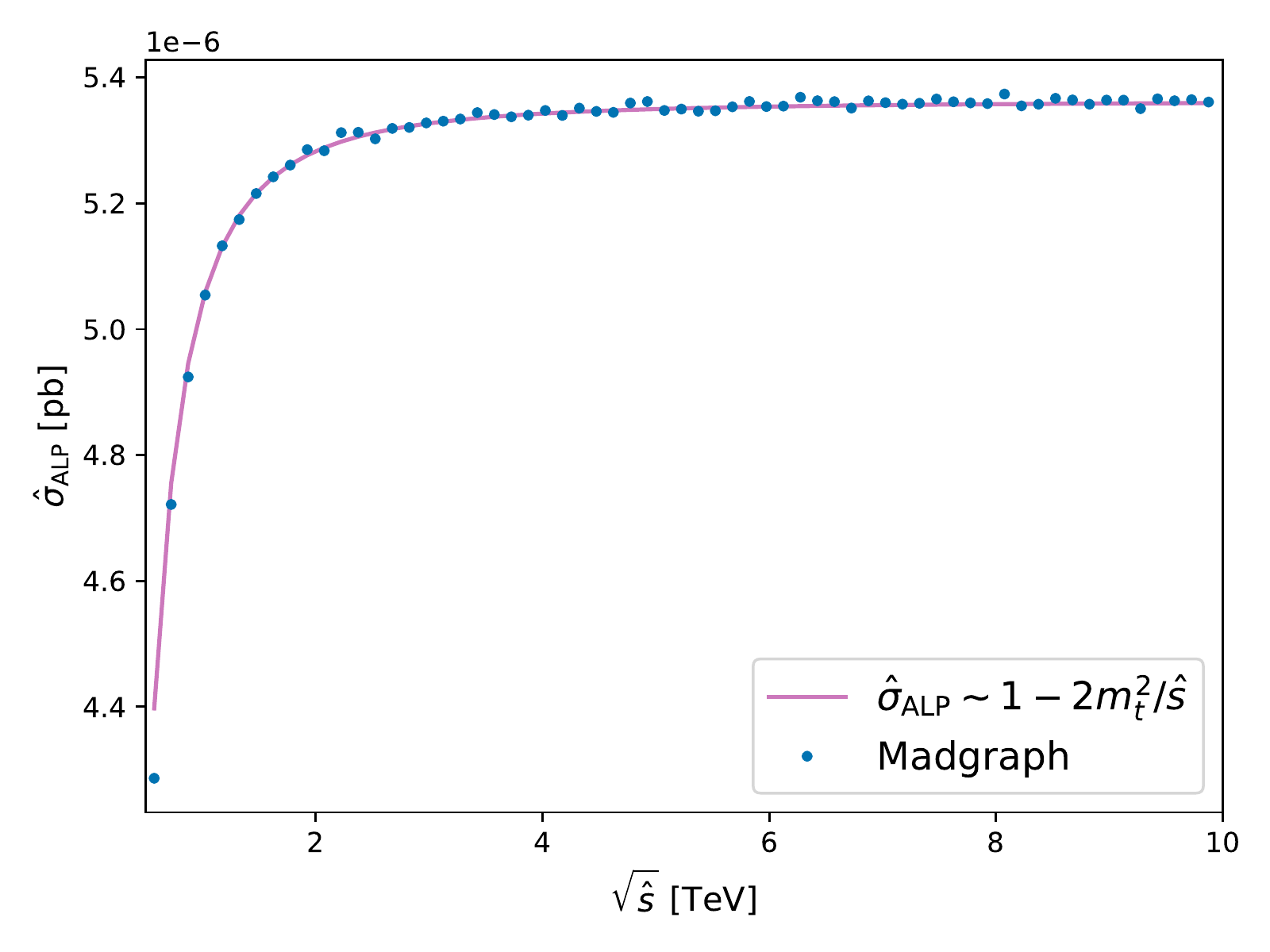}
         \caption{ALP-ALP squared contribution}
     \end{subfigure}
     \begin{subfigure}[b]{0.49\textwidth}
         \centering
         \includegraphics[width=\textwidth]{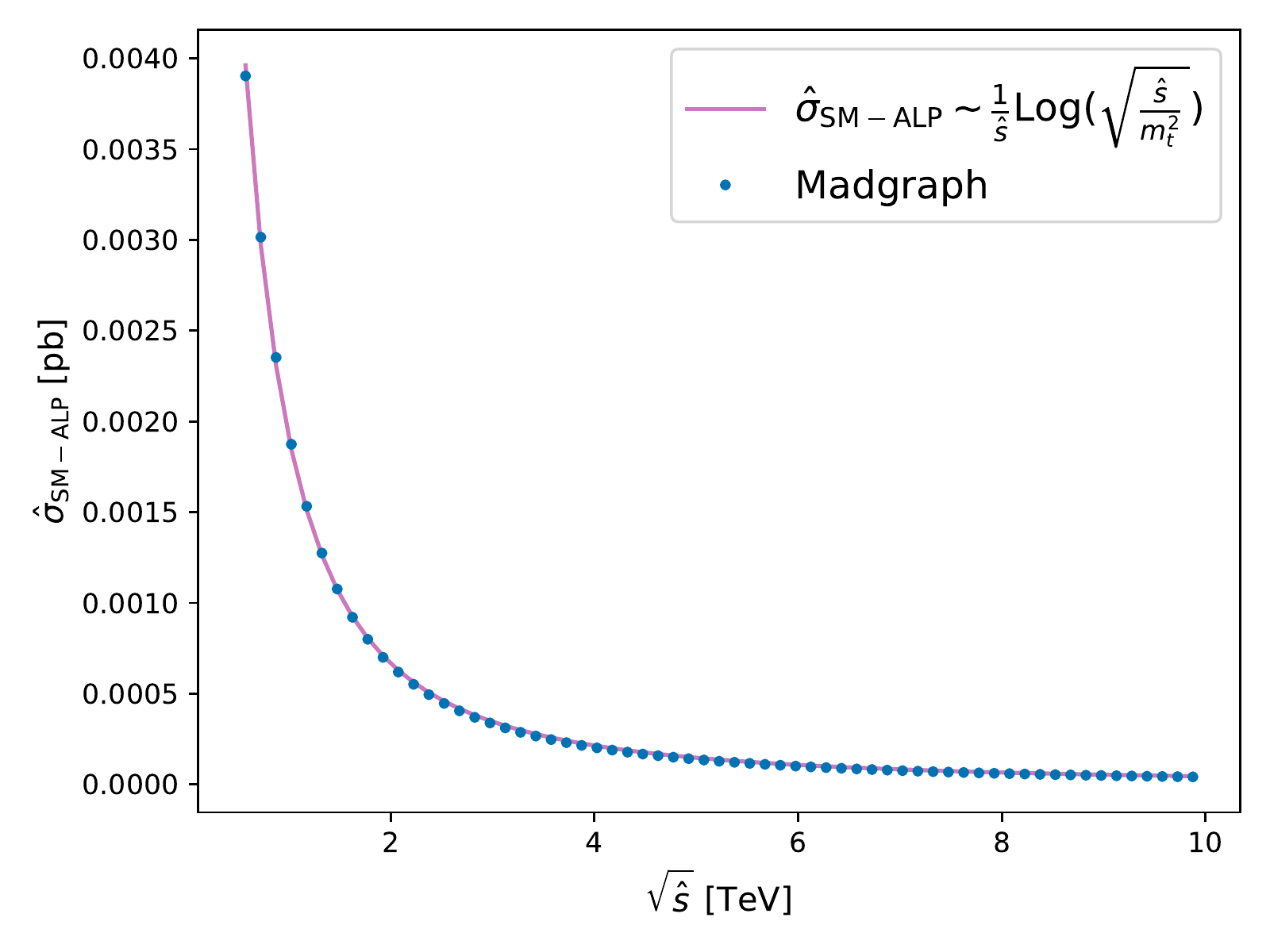}
         \caption{ALP-SM interference contribution}
     \end{subfigure}
     \begin{subfigure}[b]{0.49\textwidth}
         \centering
    \includegraphics[width=\textwidth]{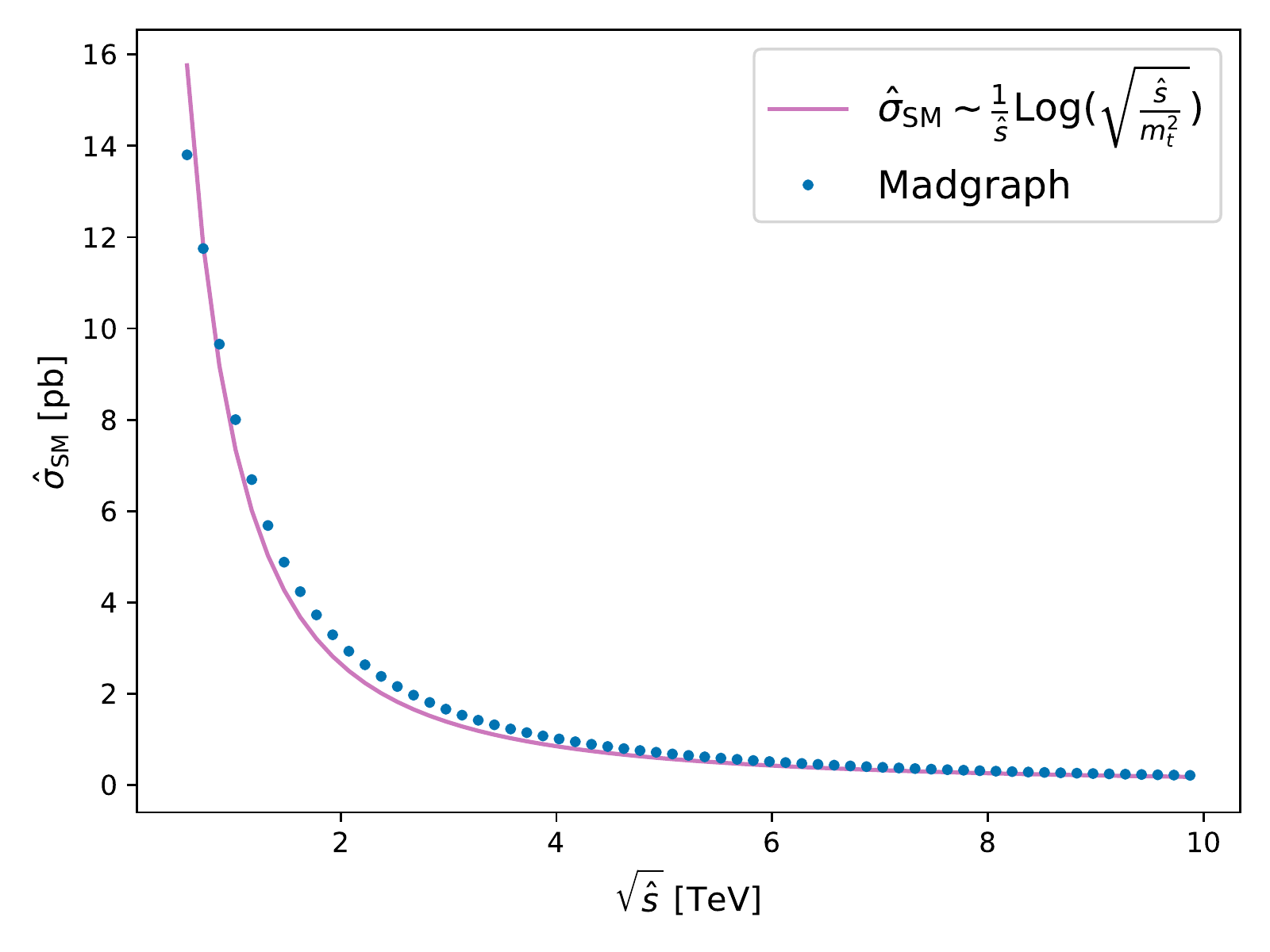}
         \caption{SM-SM gluon-channel contribution}
         \label{fig:partonic_numerical_SM}
     \end{subfigure}
     \caption{Comparison between numerical calculations of $\hat{\sigma}_{\textrm{ALP-ALP}}$,$\hat{\sigma}_{\textrm{SM-ALP}}$ and $\hat{\sigma}_{\textrm{SM-SM}}$ in the gluon channel and the large-$\hat{s}$ behaviour of their analytical expressions.}
     \label{fig:partonic_numerical}
\end{figure}

To understand the behaviour of the ALP-ALP squared contribution and SM-ALP interference contribution to the differential cross sections considered in the analysis of Sec.~\ref{sec:ALP_kinematics}, we explicitly calculate the total partonic cross section as
\begin{align}
\label{eq:sigma}
    &\hat{\sigma} = \frac{1}{2\hat{s}} \int  \frac{1}{2 (2 \pi)^2 \hat{s}} \frac{p_{3T} \, d p_{3T} \, d \theta}{\sqrt{1 - 4m_t^2/\hat{s} - 4 p_{3T}^2/\hat{s}}} |\mathcal{M}|^2 \, ,
\end{align}
where we parametrise the Lorentz invariant phase space in terms of the top quark $p_T$ and angle $\theta$ between the beam line and the emitted top quark antitop pair in the centre-of-mass frame. Substituting the ALP-ALP matrix element in Eq.~\eqref{eq:sigma}, we obtain the ALP-only contribution to the partonic cross section, which is given by
\begin{align}
    \hat{\sigma}_{\rm ALP-ALP} &\propto \frac{\alpha_s^2}{4 \pi } \frac{c_t^4}{f_a^4}  \frac{m_t^2\hat{s}^2}{(\hat{s} - m_a^2)^2}   \sqrt{1-\frac{4 m_t^2}{\hat{s}}} \\
    & \sim \frac{\alpha_s^2}{4 \pi} \frac{c_t^4}{f_a^4} \,m_t^2 \,\left(1-\frac{2 m_t^2}{\hat{s}} \right),
\end{align}
where in the second line we expanded the expression in the first line to highlight the large-$\hat{s}$ behaviour, expanding for $\hat{s} \gg m_t^2, m_a^2$.  Analogously, the contribution to the partonic cross section yielded by the SM-ALP interference is given by
\begin{align}
    \hat{\sigma}_{\rm SM-ALP} &\propto \frac{\alpha_s^2}{8 \pi} \frac{c_t^2}{ f_a^2} \frac{m_t^2}{(\hat{s} - m_a^2 )} \left( -\sqrt{1 - \frac{4 m_t^2}{\hat{s}}} + \textrm{ArcSinh}\left(\frac{\sqrt{\hat{s} - 4 m_t^2}}{2 m_t}\right) \right) \, \\
    & \sim \frac{\alpha_s^2 }{4 \pi} \frac{c_t^2}{f_a^2 } \frac{m_t^2}{\hat{s}} \log \left(\sqrt{\frac{\hat{s}}{m_t^2}} \right) \, ,
\end{align}
where again in the second line we show the large-$\hat{s}$ behaviour.  We verify our analytic calculations of $\hat{\sigma}_{\rm ALP-ALP}$ and $\hat{\sigma}_{\rm SM-ALP}$ by comparison with numerical calculations of the partonic cross sections using {\tt MadGraph5\_aMC@NLO}, as shown in Fig.~\ref{fig:partonic_numerical}.    

Note that while $\hat{\sigma}_{\rm ALP-ALP}$ tends to a constant at large-$\hat{s}$, we find $\hat{\sigma}_{\rm SM-ALP}$ decays as a function of $\hat{s}$.  
Furthermore, in some regions of phase space it decays more quickly than the gluon-gluon initiated SM contribution to the total cross section.  In particular, in the SM, the partonic cross section for $gg \rightarrow t \bar{t}$ yields~\cite{Campbell:2017hsr,Nason:1987xz}:
\begin{align}
    \hat{\sigma}_{\rm SM-SM,gg} &\propto \,\frac{\alpha_s^2}{\hat{s}^4} \Bigg( - \hat{s}^2 (31 m_t^2 + 7 \hat{s}) \sqrt{1-4 m_t^2/\hat{s}}  \\
    &\qquad + 8 \hat{s}\, (m_t^4 + 4 m_t^2 \hat{s} + \hat{s}^2) \textrm{Sinh}^{-1}\left(\frac{\sqrt{\hat{s} - 4 m_t^2}}{2 m_t}\right) \Bigg)  \notag \\
  &\sim  \frac{\alpha_s^2}{\hat{s}^2} \left( -7  \hat{s} + 8 \hat{s} \,\textrm{log}\left(\frac{\sqrt{\hat{s}}}{m_t}\right) + m_t^2 \left(-17  + 32 \, \textrm{log}\left(\frac{\sqrt{\hat{s}}}{m_t} \right) \right)  \right)
  \label{eq:SM_partonic_full}\\
  & \sim \frac{\alpha_s^2}{\hat{s}} \textrm{log}\left(\sqrt{\frac{\hat{s}}{m_t^2}} \right) \label{eq:SM_partonic_HE} \,,
\end{align}
where in the second and third passage we highlight the large-$\hat{s}$ behaviour. 
The SM $q \bar{q}$ channel contributes to the cross section with a further $s$-channel diagram.
Although at high $\hat{s}$ the SM $t$-channel dominates and $\hat{\sigma}_{\rm SM-SM}$ tends to $\frac{1}{\hat{s}} \log\big(\sqrt{\frac{\hat{s}}{m_t^2}} \big)$, exactly like $\hat{\sigma}_{\rm SM-ALP}$, we find that, even at $\hat{s} \sim 1-2$ TeV, the subleading $\hat{s}^{-2}$ contribution to $\hat{\sigma}_{\rm SM-SM}$ is subdominant but non-negligible.  We can see this explicitly in Fig.~\ref{fig:partonic_numerical_SM}, which demonstrates that Eq.~\ref{eq:SM_partonic_HE} provides a perfect description of $\hat{\sigma}_{\textrm{SM-SM}}$ only at very high $\sqrt{\hat{s}} \gtrsim 4$ TeV, while it is approximated at lower values of $\sqrt{\hat{s}}$.
\begin{figure}[htb!]
    \centering
    \includegraphics[scale=0.7]{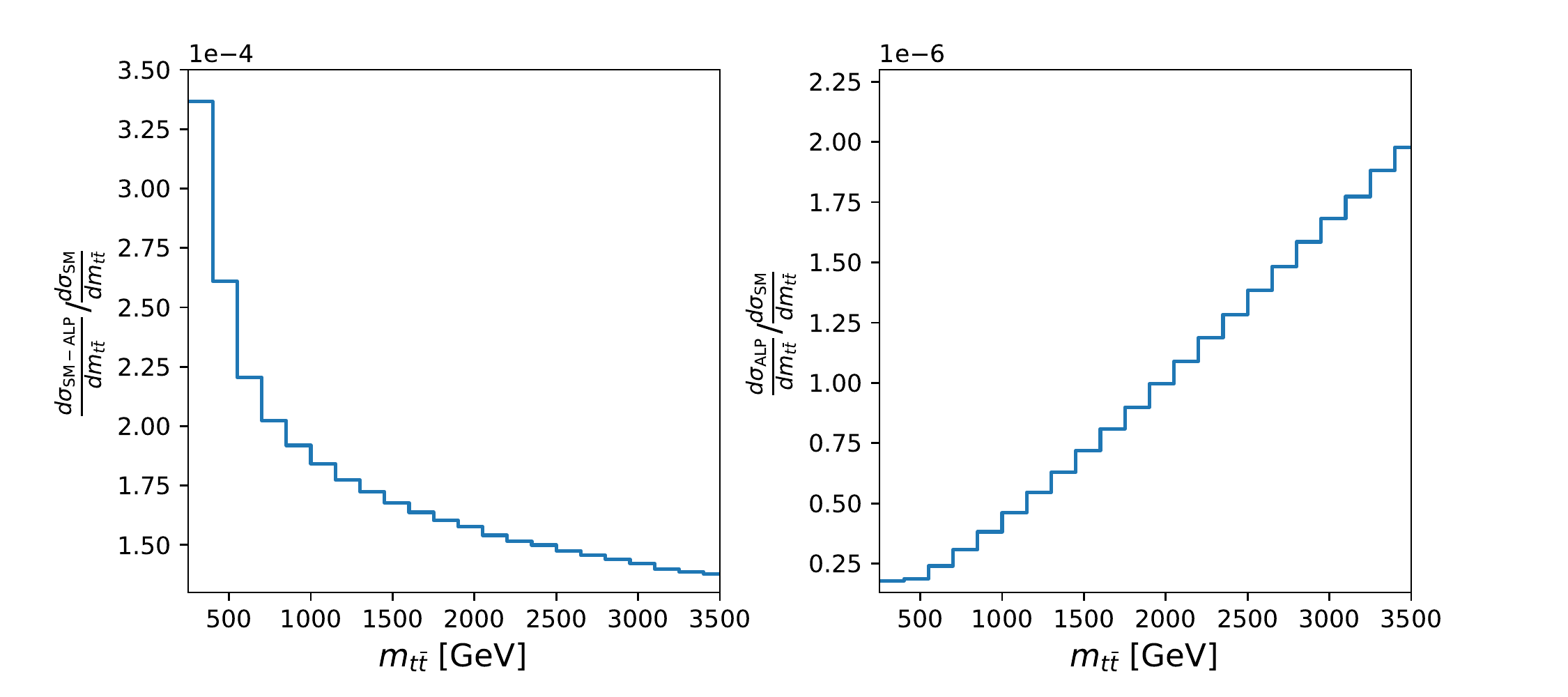}
    \caption{Ratio of the differential cross sections resulting from ALP-SM interference (left) and the ALP-ALP squared contribution (right) to the SM differential cross section as a function of $m_{t \bar{t}}$, assuming $c_t/f_a=1~{\rm TeV}^{-1}$. Here we plot the total cross section and we use as input PDFs the {\tt NNPDF4.0 NNLO} set with $\alpha_s(M_Z)=0.118$~\cite{NNPDF:2021njg}.}
    \label{fig:APP-ALPSMkinematics}
\end{figure}

As a result, we observe that the SM-ALP differential cross section decays with $m_{t \bar{t}}$ relative to the SM in the kinematic region of interest for the analysis considered in Sec.~\ref{sec:indirect}.  This is shown at the level of the total cross sections in Fig.~\ref{fig:APP-ALPSMkinematics}, after the convolution of the partonic cross sections with the PDFs.  There we display the differential cross section resulting from ALP-SM interference and from the ALP-ALP squared contribution, both normalised to the SM differential cross section as a function of $m_{t \bar{t}}$.
It can be seen that $\frac{d \sigma_{\textrm{ALP-ALP}}}{d m_{t \bar{t}}}$ grows with $m_{t \bar{t}}$ relative to $\frac{d \sigma_{\textrm{SM-SM}}}{d m_{t \bar{t}}}$, despite the fact that the gluon PDF dampens the $\hat{s}$ growth observed in $\hat{\sigma}_{\textrm{ALP-ALP}}$.  Plots are shown for $c_t/f_a=1~{\rm TeV}^{-1}$.

\begin{figure}[htb!]
     \centering
     \begin{subfigure}[b]{0.49\textwidth}
         \centering
         \includegraphics[width=\textwidth]{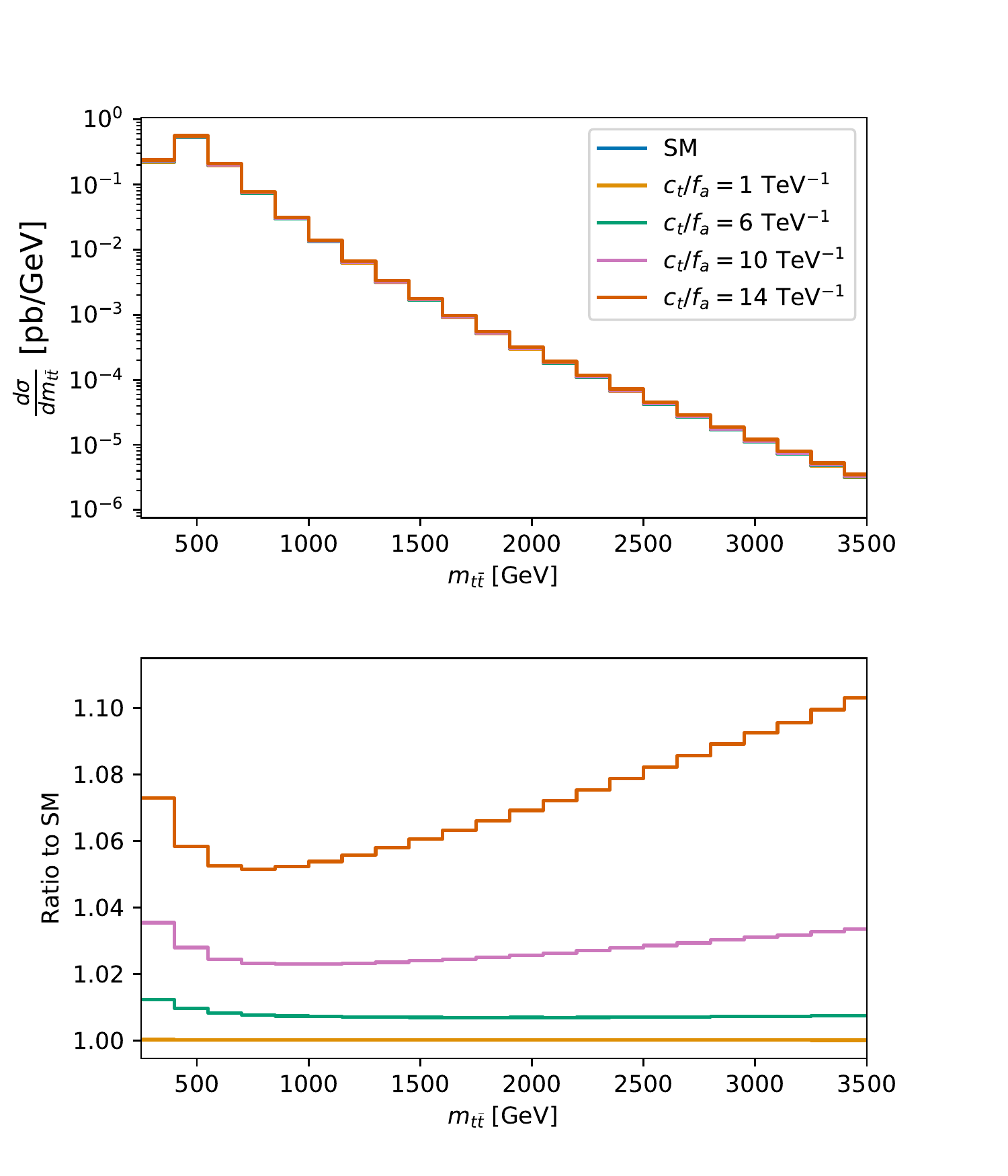}
         \caption{}
     \end{subfigure}
     \begin{subfigure}[b]{0.49\textwidth}
         \centering
         \includegraphics[width=\textwidth]{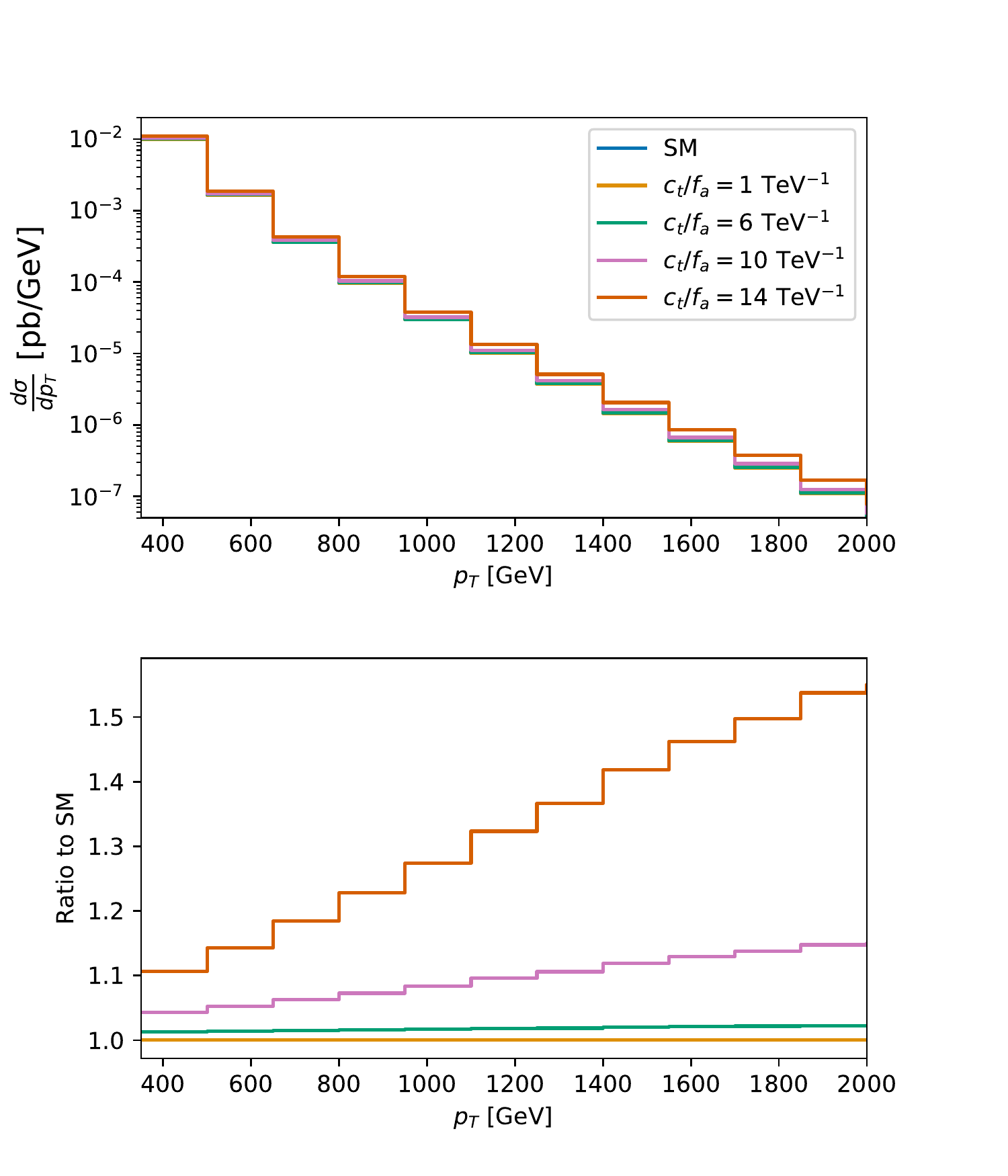}
         \caption{}
     \end{subfigure}
     \caption{Contribution of the ALP signal to $\frac{d \sigma}{d m_{t \bar{t}}}$ (left) and $\frac{d \sigma}{d p_{T}^t}$ (right), evaluated at the ALP-top couplings $(c_t/f_a) = 1~{\rm TeV}^{-1}$ shown. The input PDF set is the {\tt NNPDF4.0 NNLO} set with $\alpha_s(M_Z)=0.118$~\cite{NNPDF:2021njg}.}
    \label{fig:APP-ALPSMsignal}
\end{figure}

In Fig.~\ref{fig:APP-ALPSMsignal} we explore the combined effect of the ALP-ALP and ALP-SM interference for various values of $(c_t/f_a)$.  Here we see explicitly that the effect of the SM-ALP interference term dominates over the ALP-ALP signal for $c_t < 10$ and counteracts the growth with $m_{t \bar{t}}$ provided by it. Larger values of $c_t$ are needed for the squared ALP signal term to dominate over the interference term.
Finally, in the same figure, Fig.~\ref{fig:APP-ALPSMsignal} (b), we show the effect of the ALP-ALP signal on the $p_T$ spectrum of the top quark, assuming a boosted top quark with $p_T > 355$ GeV, as measured by ATLAS~\cite{ATLAS:2022xfj}, for example.  Similarly to the $m_{t \bar{t}}$ spectrum, the ratio of the interference term to the SM decays with $p_{T}$.  However, the interference term is suppressed by working in the high-$p_{T}$ regime, and we observe that the ALP-ALP signal produces a growth with $p_{T}$ in the distribution tail, further motivating the use of this $p_{T}$ spectrum in constraining the ALP-top coupling in Sec.~\ref{sec:indirect}.

\bibliographystyle{JHEP}
\bibliography{axiontop}

\end{document}